\documentclass[aps,preprint,superscriptaddress]{revtex4-2}

\usepackage{amsmath}
\usepackage{graphicx}
\usepackage{bm}

\preprint{APS/123-QED}
\begin{document}
	
	\noindent
	
	\title[Email of corresponding author: mjgerard2@wisc.edu]{On the effect of flux-surface shaping on trapped-electron modes in quasi-helically symmetric stellarators}
	
	\author{M.J.~Gerard \footnote{Corresponding author: mjgerard2@wisc.edu}}
	
	\affiliation{University of Wisconsin-Madison, Wisconsin 53706, USA}
	
	\author{M.J.~Pueschel}
	
	\affiliation{Dutch Institute for Fundamental Research, 5612 AJ Eindhoven, The Netherlands}
	\affiliation{Eindhoven University of Technology, 5600 MB Eindhoven, The Netherlands}
	\affiliation{Department of Physics \& Astronomy, Ruhr-Universit\"at Bochum, 44780 Bochum, Germany}
	
	\author{B.~Geiger}
	
	\affiliation{University of Wisconsin-Madison, Wisconsin 53706, USA}
	
	\author{R.J.J.~Mackenbach}
	
	\affiliation{Eindhoven University of Technology, 5600 MB Eindhoven, The Netherlands}
	\affiliation{Max Planck Institute for Plasma Physics, 17491 Greifswald, Germany}
	\affiliation{Ecole Polytechnique F\'ed\'erale de Lausanne, Swiss Plasma Center, CH-1015 Lausanne, Switzerland}
	
	\author{J.M.~Duff}
	
	\affiliation{University of Wisconsin-Madison, Wisconsin 53706, USA}
	
	\author{B.J.~Faber}
	\author{C.C.~Hegna}
	\author{P.W.~Terry}
	
	\affiliation{University of Wisconsin-Madison, Wisconsin 53706, USA}
	
	\begin{abstract}
		Using a novel optimization procedure it has been shown that the Helically Symmetric eXperiment (HSX) stellarator can be optimized for reduced trapped-electron-mode (TEM) instability [M.J.~Gerard et al., \textit{Nucl.~Fusion} \textbf{63} (2023) 056004]. Presently, with a set of 563 experimental candidate configurations, gyrokinetic simulations are performed to investigate the efficacy of available energy $E_\mathrm{A}$, quasi-helical symmetry, and flux-surface shaping parameters as metrics for TEM stabilization. It is found that lower values of $E_\mathrm{A}$ correlate with reduced growth rates, but only when separate flux-surface shaping regimes are considered. Moreover, configurations with improved quasi-helical symmetry demonstrate a similar reduction in growth rates and less scatter compared to $E_\mathrm{A}$. Regarding flux-surface shaping, a set of helical shaping parameters is introduced that show increased elongation is strongly correlated with reduced TEM growth rates, however, only when the quasi-helical symmetry is preserved. Using a newly derived velocity-space-averaged TEM resonance operator, these trends are analyzed to provide insights into the physical mechanism of the observed stabilization. For elongation, stabilization is attributed to geometric effects that reduce the destabilizing particle drifts across the magnetic field. Regarding quasi-helical symmetry, the TEM resonance in the maximally resonant trapping well is shown to increase as the quasi-helical symmetry is broken, and breaking quasi-helical symmetry increases the prevalence of highly resonant trapping wells. While these results demonstrate the limitations of using any single metric as a linear TEM proxy, it is shown that quasi-helical symmetry and plasma elongation are highly effective metrics for reducing TEM growth rates in helical equilibria.
	\end{abstract}
	
	\maketitle
	
	\section{Introduction \label{sec:introduction}}
	
	Reaching burning plasma conditions in magnetic confinement fusion (MCF) devices requires high densities, high temperatures, and long energy-confinement times. However, most fusion devices show a strong degradation of the confinement time during high-temperature and high-density operation. This reduction can be attributed primarily to the cross-field transport arising from turbulence at the gyroradius scale, which is driven by microinstabilities occurring at comparable scales \cite{yamada_characterization_2005}.
	
	One technique for reducing turbulent transport is to modify the magnetic field geometry \cite{mynick_optimizing_2010, xanthopoulos_controlling_2014, proll_tem_2016, hegna_theory_2018, Terry_2018, Nunami_2022, roberg-clark_2022, kim2024optimization}. In the stellarator approach to fusion, this can be achieved with the coils used to generate the magnetic field. The first stellarator in the world to demonstrate the efficacy of magnetic-field shaping for reducing neoclassical transport is the Helically Symmetric eXperiment (HSX) \cite{canik_experimental_2007, canik_reduced_2007}. HSX is a four-fold symmetric device that can generate a plasma with approximately $1.2$ m major radius, $0.12$ m minor radius, and plasma volume around $0.44$ m$^3$. It was designed to be able to generate different magnetic topologies by varying the current in its external coils. Notably, the device can generate configurations with both quasi-helical symmetry and broken quasi-helical symmetry, though the standard configuration, referred to as the QHS configuration, is quasi-helically symmetric \cite{almagri_helically_1999, Talmadge_PoP_2001}. Due to the successful neoclassical optimization of HSX, the dominant source of transport is attributed to turbulence. With densities around $10^{18} \ \mathrm{m}^{-3}$ and electron-cyclotron heating, the electrons in HSX are much hotter than the ions. Moreover, peaked density profiles are typically observed \cite{kumar_radial_2018}. Therefore, the ion-temperature-gradient (ITG) mode tends not to contribute significantly to the cross-field transport, though ITG modes in HSX geometries have been studied computationally \cite{McKinney_JPP_2019, McKinney_JPP_2021}. Alternatively, simulation work has shown that the trapped-electron mode (TEM) is destabilized in HSX, and likely drives the anomalous transport in the device \cite{guttenfelder_effect_2008, faber_gyrokinetic_2015}.
	
	Recently, it has been demonstrated computationally that the collisionless TEM growth rates can be reduced in HSX by elongating the plasma while preserving the quasi-helical symmetry \cite{Gerard_2023}. Those results are based on a set of 100 configurations selected from a database of over 1 million unique magnetohydrodynamic (MHD) HSX-based equilibria. This database was produced by numerically modifying coil currents in the 48 auxiliary and 48 main coils of the experiment. This flexibility in HSX provides a useful domain for testing the efficacy of different TEM optimization metrics while simultaneously investigating the relationship between the magnetic field geometry and the governing TEM resonance. Therefore, in the present paper, the number of configurations investigated in linear gyrokinetic simulation is increased to include configurations with broken quasi-helical symmetry. Moreover, the simulation results are compared against several TEM metrics, and the physical mechanism governing the observed trends is discussed. This process, while consistent with results from Ref.~\cite{Gerard_2023}, reveals new insights into the optimization of quasi-helically symmetric equilibria for reduced TEM growth rates.
	
	The collisionless TEM is destabilized by a resonance between the precessional drift of the trapped electrons and the electrostatic drift wave. This has been reported extensively for tokamak geometries \cite{kadomtsev_trapped_1971, adam_destabilization_1976, catto_trapped_1978} and more recently has been shown to apply to stellarators \cite{proll_resilience_2012, helander_collisionless_2013, proll_collisionless_2013}. The stability condition for TEMs is well characterized by the equation for the energy transfer rate between the trapped electrons and the electrostatic drift wave \cite{proll_resilience_2012, helander_collisionless_2013, proll_collisionless_2013, proll_tem_2016}. Near marginal stability ($\gamma \rightarrow 0^+$), this equation can be expressed as
	\begin{equation}
		P_\mathrm{e} = \frac{\pi e^2}{T_\mathrm{e0}} \int_{-\infty}^{\infty} \frac{d\ell}{B} \int_{-\infty}^{\infty} \omega \left( \omega - \omega_\mathrm{*e}^T \right) \delta\left( \omega - \overline{\omega}_\mathrm{de} \right) |\overline{\phi}|^2 f_\mathrm{e0} d^3v, \label{eq:power_transfer}
	\end{equation}
	which was first derived in Ref.~\cite{proll_resilience_2012}. Throughout this paper, an overbar is used to denote a bounce average, which is defined in Eq.~(\ref{eq:bounce-average}). Then, $e$ is the elementary charge, $T_\mathrm{e0}$ the background electron temperature, $\ell$ the magnetic-field-line arc length, $B$ the magnetic-field strength, $\overline{\phi}$ the bounce-averaged electrostatic potential, $f_\mathrm{e0}$ the zeroth-order electron distribution-function, $v$ the particle velocity, and $\delta$ the Dirac-delta distribution-function. There are also three frequencies in the formula, of which $\omega$ is the electrostatic drift-wave frequency, $\overline{\omega}_\mathrm{de}$ is the electron bounce-averaged-drift frequency, and $\omega_\mathrm{*e}^T$ the electron-diamagnetic-drift frequency. Note that the $e$ subscript will be dropped for the remainder of this paper.
	
	The diamagnetic-drift frequency is defined as
	\begin{equation}
		\omega_\mathrm{*}^T = \frac{k_{\alpha}T_\mathrm{0}}{e} \left[ \frac{\partial \ln n_\mathrm{0}}{\partial\psi} + \frac{\partial \ln T_\mathrm{0}}{\partial\psi} \left( \frac{E}{T_\mathrm{0}} - \frac{3}{2} \right) \right], \label{eq:diamagnetic}
	\end{equation}
	where $n_\mathrm{0}$ denotes the background electron density, $E=mv^2/2$ is the kinetic energy with electron mass $m$ and velocity magnitude $v$, $\psi$ is the toroidal magnetic flux, and $\alpha$ is the Clebsh angle, which locally defines the magnetic field as $\mathbf{B} = \nabla\psi \times \nabla\alpha$. Then, $k_{\alpha}$ is the wavenumber in the $\nabla\alpha$ direction. The bounce-averaged-drift frequency is defined as
	\begin{equation}
		\overline{\omega}_\mathrm{d} = k_{\alpha} \overline{\mathbf{v}_\mathrm{d} \cdot \nabla\alpha} + k_{\psi} \overline{\mathbf{v}_\mathrm{d} \cdot \nabla\psi}, \label{eq:drift_frequency}
	\end{equation}
	where $\mathbf{v}_\mathrm{d}$ is the electron-drift velocity and $k_{\psi}$ is the wavenumber in the $\nabla\psi$ direction. The two components of the bounce-averaged-drift frequency are further identified as the bounce-averaged precessional-drift frequency $\overline{\omega}_{\alpha} = k_{\alpha} \overline{\mathbf{v}_\mathrm{d} \cdot \nabla\alpha}$ and the bounce-averaged radial-drift frequency $\overline{\omega}_{\psi} = k_{\psi} \overline{\mathbf{v}_\mathrm{d} \cdot \nabla\psi}$. In terms of a particle's second adiabatic invariant $\mathcal{J} = \int mv_{\parallel} d\ell$, with $v_{\parallel}$ its parallel velocity, the precessional- and radial-drift frequencies can be expressed as
	\begin{align}
		\overline{\omega}_{\alpha} &= -\frac{k_{\alpha}}{e} \left( \frac{\partial \mathcal{J}}{\partial \psi} \right)_{E,\mu, \alpha} \bigg/ \left( \frac{\partial \mathcal{J}}{\partial E} \right)_{\mu,\psi,\alpha}, \label{eq:precessional_drift} \\
		\overline{\omega}_{\psi} &= \frac{k_{\psi}}{e} \left( \frac{\partial\mathcal{J}}{\partial\alpha} \right)_{E,\mu,\psi} \bigg/ \left( \frac{\partial\mathcal{J}}{\partial E} \right)_{\mu,\psi,\alpha} \label{eq:radial_drift}
	\end{align}
	where the subscripts denote variables being held constant, with $\mu = mv_{\perp}^2/2B$ the magnetic moment and $v_{\perp}$ the magnitude of the particle's perpendicular velocity, relative to the magnetic field.
	
	When $P < 0$, Eq.~(\ref{eq:power_transfer}) describes a destabilizing transfer of energy from the trapped electrons into the drift wave via resonant interaction. Therefore, the destabilization of a TEM requires that $\omega (\omega - \omega_*^T)$, or equivalently $\overline{\omega}_\mathrm{d}(\overline{\omega}_\mathrm{d} - \omega_*^T)$, be negative in regions along a magnetic field line where the trapped-particle fraction is high and the bounce-averaged electrostatic eigenfuction is peaked. The other terms in the expression then provide a flux-surface and velocity-space weighting of this condition. Note that in so-called omnigenous magnetic fields, where the bounce-averaged radial drifts of trapped particles goes to zero ($\overline{\omega}_{\psi} = 0$), then $\overline{\omega}_\mathrm{d} = \overline{\omega}_{\alpha}$. Therefore, to produce an unstable TEM, an omnigenous magnetic field requires that the diamagnetic drift and the bounce-averaged precessional drift propagate in the same direction. To be sure, no magnetic field is perfectly omnigenous; however, quasi-omnigenous configurations with $\overline{\omega}_{\psi}/\overline{\omega}_{\alpha} \ll 1$ do exist, in which case the destabilization of TEMs is due to unfavorable precessional drifts in the electron-diamagnetic direction. Since the QHS configuration is quasi-helically symmetric, and therefore quasi-omnigenous, this is also the case for the collisionless TEM in the QHS configuration.
	
	Calculation of the energy transfer described in Eq.~(\ref{eq:power_transfer}) requires knowledge of the drift-wave frequency and the bounce-averaged electrostatic potential, both of which typically require explicit calculation as done in the gyrokinetic code \textsc{Gene} \cite{noauthor_gene_nodate, jenko_electron_2000}. For stellarator optimization, the more computationally efficient calculation of the available energy $E_\mathrm{A}$ could be used as a proxy for TEM growth rates. $E_\mathrm{A}$ is a TEM metric that defines an upper bound on the amount of energy that is available in the trapped-electron population to drive fluctuations \cite{helander_2017, helander_2020, Mackenbach_2022, Mackenbach_2023_JPP}. As described in Ref.~\cite{Mackenbach_2022}, $E_\mathrm{A}$ can be calculated in flux-tube geometry as
	\begin{equation}
		E_\mathrm{A} = \pi^2 \left( \frac{e\Delta\psi \Delta\alpha}{m} \right)^2 \int\int \frac{f_0}{T_\mathrm{e0}} \left[ \overline{\omega}_{\alpha}^2 \left( \frac{\omega_*^T}{\overline{\omega}_{\alpha}} - 1 + F \right) \frac{\Delta\psi}{\Delta\alpha} + \overline{\omega}_{\psi}^2 \left(-1 + F\right) \frac{\Delta\alpha}{\Delta\psi} \right] d\mu d\mathcal{J}, \label{eq:aval}
	\end{equation}
	where $\Delta\psi$ and $\Delta\alpha$ are the length scales over which energy is available and
	\begin{equation}
		F = \sqrt{\frac{(\omega_*^T - \overline{\omega}_{\alpha})^2 (\Delta\psi)^2 + \overline{\omega}_{\psi}^2 (\Delta\alpha)^2}{ \overline{\omega}_{\alpha}^2 (\Delta\psi)^2 + \overline{\omega}_{\psi}^2 (\Delta\alpha)^2 }}. \label{eq:F_ae}
	\end{equation}
	Note that the flux-tube $E_\mathrm{A}$ can be normalized as
	\begin{equation}
		\hat{E}_\mathrm{A} = \frac{E_\mathrm{A}}{E_\mathrm{th}\rho_*^2}, \label{eq:aval_norm}
	\end{equation}
	where $\rho_* = \rho_\mathrm{s}/a$ with $\rho_\mathrm{s}$ the ion gyroradius at the ion sound speed, $a$ is the effective-minor-radius at the last closed flux surface (LCFS), and
	\begin{equation}
		E_\mathrm{th} = \frac{3}{2}n_0T_0 \pi \Delta\psi  \Delta\alpha \int \frac{d\ell}{B}
	\end{equation}
	is the thermal energy of a plasma in a flux tube to leading order in an expansion in the directions perpendicular to the magnetic field \cite{Mackenbach_2023_JPP}.
	
	It is important to note the significance of $\Delta\psi$ and $\Delta\alpha$ in the calculation of $E_\mathrm{A}$. In a gyrokinetic simulation, one may resolve fluctuation dynamics in a volumetrically minimized domain by making a flux-tube approximation \cite{beer_1995}. Importantly, in this local approximation, increasing the simulation domain beyond this minimally resolved domain does not increase fluctuation energies. Therefore, for $E_\mathrm{A}$ to relate to these local fluctuations, $\Delta\psi$ and $\Delta\alpha$ must correspond to the length scale of those fluctuations. Following arguments presented in Ref.~\cite{Mackenbach_2023_JPP}, these free parameters in the available energy are defined as $\Delta\psi = \rho_*^2\psi_\mathrm{edge}$ and $\Delta\alpha = \rho_*/\sqrt{s}$, where $s = \psi/\psi_\mathrm{edge}$ and $\psi_\mathrm{edge}$ is the toroidal magnetic-flux at the LCFS, so that $s$ labels the flux surface being considered.
	
	The concept of available energy has been used to argue for a possible connection between reduced TEM turbulence and reduced neoclassical radial particle transport in the collisionless regime. This conjecture is motivated by the observation that a Maxwellian distribution can only satisfy the minimal-$E_\mathrm{A}$ state if the magnetic field is omnigenous \cite{helander_2020, Mackenbach_2022}. This is observed in Eq.~(\ref{eq:aval}) as a monotonically increasing dependence of $E_\mathrm{A}$ on $\overline{\omega}^2_{\psi}$. Since quasi-omnigeneity is achieved in HSX through quasi-helical symmetry, TEM growth rates will be compared against the $E_\mathrm{A}$ and the quasi-helical symmetry of various HSX equilibria.
	
	Quasi-helical symmetry can be quantified as
	\begin{equation}
		\mathcal{Q} = \frac{1}{B_{0,0}} \left( \sum\limits_{m = 0}^{M} \sum\limits_{\substack{n=-4N\\n\neq4m}}^{4N} B^2_{n,m} \right)^{1/2}, \label{eq:symmetry}
	\end{equation}
	where $N$ and $M$ are the number of toroidal and poloidal modes in a straight field-line coordinate system and $B_{n,m}$ are the mode amplitudes of the magnetic field strength on a flux surface \cite{boozer_guiding_1980}. Due to the $n=4m$ symmetry in the QHS magnetic field, $\mathcal{Q}$ quantifies the extent to which the symmetry is broken in the magnetic field's so-called Boozer spectrum.    
	
	Recent analytic work has also shown the resiliency of so-called maximum-$\mathcal{J}$ configurations towards TEM stability \cite{proll_resilience_2012, helander_collisionless_2013}, where maximum-$\mathcal{J}$ means that $\partial\mathcal{J}/\partial\psi < 0$ throughout an equilibrium. Notably, this property is locally satisfied for deeply trapped particles in quasi-isodynamic configurations like Wendelstein 7-X (W7-X) and is achieved over a broader set of trapped-electron pitch-angles in new configurations presented in Refs.~\cite{Sanchez_2023, rodriguez_2023}. The effect of this maximum-$\mathcal{J}$ condition is to reduce the TEM resonance by confining the electron distribution function to a domain in which trapped electrons dominantly precess in the ion-diamagnetic direction \cite{proll_resilience_2012}. Despite the absence of TEMs, gyrokinetic simulations in W7-X geometries still show finite growth rates for modes propagating the electron-diamagnetic direction \cite{costello_2023}. These modes have been identified as the universal instability (UI), which can be destabilized by the passing- or trapped-electron population, and these modes dominantly appear in situations when the TEM has been sufficiently stabilized \cite{helander_2015}.
	
	In this paper, a set of 563 unique HSX equilibria that span a broad range of coil-current perturbations is analyzed for their TEM growth rates. This allows for the testing of existing TEM metrics and analytic theories. Section~\ref{sec:selection} provides details on the set of equilibria that have been selected, with a description of how $E_\mathrm{A}$ and quasi-helical symmetry can be used in conjunction with a set of helically defined flux-surface shaping parameters. In Sec.~\ref{sec:gyrokinetic}, results from linear gyrokinetic simulations are shown to be consistent with findings in Ref.~\cite{Gerard_2023}, where the lowest growth rates were observed in highly elongated configurations. However, it is observed here that elongation must not come at the expense of quasi-helical symmetry, which, when broken, leads to an increase in growth rates. Moreover, no universal correlation between TEM growth rates and $E_\mathrm{A}$ is observed. Instead, both $E_\mathrm{A}$ and quasi-helical symmetry are shown to scale with growth rates when considered in particular shaping regimes. A velocity-space-averaged TEM resonance operator is defined in Sec.~\ref{sec:resonance} and compared against the growth rates. From this analysis, it is found that the reduction in growth rates is consistent with the energy transfer rate defined in Eq.~(\ref{eq:power_transfer}), but that the stabilization does not result in a transition from a dominant TEM to UI. By analyzing trapped-electron drifts in helically linked magnetic trapping wells, the growth-rate reduction with an increase in elongation is shown to be the result of an increase in the rotational transform and a modification in the magnetic-field-strength spectrum that preserves quasi-helical symmetry. Moreover, the dependence of growth rates on quasi-helical symmetry is shown to be caused by an increase in the prevalence of highly resonant trapping wells. These results show that elongation can be used along with either $E_\mathrm{A}$ or a quasi-helical-symmetry metric as a promising combination of metrics for the optimization of helical equilibria for reduced TEM growth rates. Moreover, this further demonstrates the efficacy of the optimization technique introduced in Ref.~\cite{Gerard_2023} of exploring the configuration space surrounding the operating point of an existing experiment.	
	
	\section{Equilibrium Selection \label{sec:selection}}
	
	\subsection{Selection Metrics \label{sec:selection_metrics}}
	
\begin{figure}
		\centering
		\includegraphics[width=.48\textwidth, keepaspectratio]{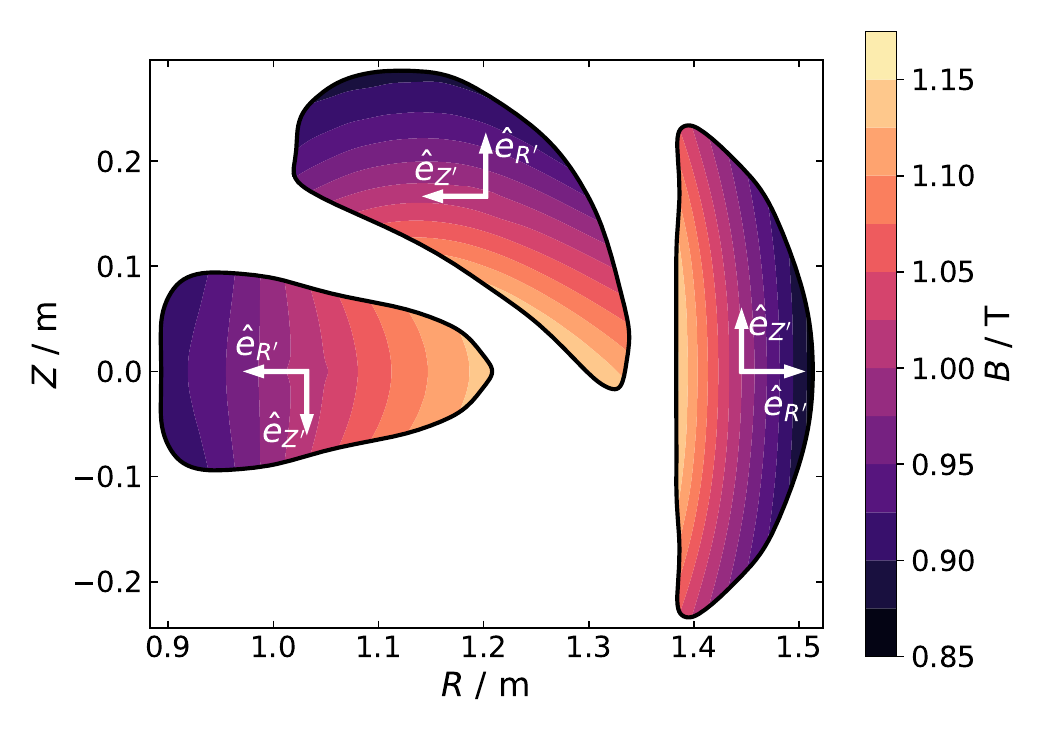}
		\caption{Three toroidal cross-sections are shown for the QHS equilibrium, with the helically rotated basis vectors shown after translation to the magnetic axis. It is within this reference frame that the helical shaping parameters are calculated. Importantly, $\hat{\mathbf{e}}_{R^{\prime}}$ is constructed to point towards the low-field side of a given cross-section. \label{fig:helical_frame}}
	\end{figure}

	In this section, a set of new shaping parameters is defined for helical equilibria, which will be used to characterize flux-surface geometries throughout the HSX coil-current database. The shaping parameters are based on those defined in Ref.~\cite{Luce_2013} for an axisymmetric system. Therein, definitions are provided for flux-surface elongation $\kappa$, triangularity $\delta$, and squareness $\zeta$, all of which are based on the extrema along a flux surface at a single toroidal angle in a cylindrical coordinate system. The primary adaptation made to extend these definitions to a non-axisymmetric system is to define a rotated reference frame from which the flux-surface shapes can be calculated. The rotated reference frame is defined as
	\begin{equation}
		\begin{pmatrix}
			R^{\prime} \\
			Z^{\prime}
		\end{pmatrix}
		=
		\begin{pmatrix}
			\cos(N_\mathrm{fp}\varphi) & \sin(N_\mathrm{fp}\varphi) \\
			-\sin(N_\mathrm{fp}\varphi) & \cos(N_\mathrm{fp}\varphi)
		\end{pmatrix}
		\begin{pmatrix}
			R \\
			Z
		\end{pmatrix}, \label{eq:rotation}
	\end{equation}
	where $\{R,\, Z\}$ and $\{R^{\prime},\, Z^{\prime}\}$ are the radial and vertical coordinates in the cylindrical and rotated reference frames, respectively, and $N_\mathrm{fp}$ and $\varphi$ are the number of field periods and toroidal angle, respectively. This rotated reference frame is selected so that, relative to the magnetic axis, the radial-like direction $\hat{\mathbf{e}}_{R^{\prime}} = \nabla R^{\prime} / |\nabla R^{\prime}|$ always points towards the low-field side of a toroidal cross-section, making the shaping parameters analogous to those of the axisymmetric system. Figure~\ref{fig:helical_frame} demonstrates this rotated reference frame in three toroidal cross-sections of the QHS equilibrium, with the magnetic field strength shown in color and the rotated basis vectors shown after translation to the magnetic axis.
	
	It should be noted that in Ref.~\cite{Luce_2013}, a distinction is made between upper and lower values of elongation and triangularity. For squareness, a further distinction between upper/lower and inner/outer is made. For our purposes, all such distinctions are calculated and then averaged together at each toroidal angle. Moreover, due to the lack of axisymmetry, each shaping parameter is a function of toroidal angle. Thus, an integrated toroidal average for each shaping parameter is used as the final parameter under consideration. Though helpful in characterizing flux-surface shapes, it should be realized that because of this toroidal averaging, these shaping parameters do not uniquely describe a flux-surface geometry.
	
	\subsection{Selection Process \label{sec:selection_process}}
	
	An initial set of 600 configurations, of which 563 will be retained for gyrokinetic analysis, is selected from the database of $> 10^6$ MHD equilibria, with the database described in Ref.~\cite{Gerard_2023}. These configurations are selected using the helical shaping and symmetry metrics. To ensure all metrics are calculated along flux surfaces with the same real-space position, different flux-surface labels are selected for consideration in each configuration. With flux-surface label $s = \psi/\psi_{\mathrm{edge}} = (r/a)^2$, where $r$ is the effective-minor-radius of the flux surface, the standard flux-surface is taken to be the QHS surface at $s_\mathrm{q}=0.5$. For each $i^{\mathrm{th}}$ configuration it is then required that $r_i = r_\mathrm{q}$. This leads to the salient relation $s_i = (a_\mathrm{q}/a_i)^2 s_\mathrm{q}$, where $s_i$ labels each flux surface to be considered.
	
	The three shaping parameters, the quasi-helical symmetry metric, and normalized flux-tube $E_\mathrm{A}$ are all calculated along the $s_i$ flux surfaces and normalized to their corresponding quantity on the $s_\mathrm{q}=0.5$ flux surface, with the latter denoted by a star superscript. Since both triangularity and squareness can take on either positive or negative values, the absolute value of the corresponding metric in the QHS configuration is used in the normalization. In this way, the sign of the normalized metrics is determined by the configuration being considered. For the $E_\mathrm{A}$ calculation, the $\alpha = 0$ field line is selected and followed for four poloidal turns with 512 grid points along the direction of the magnetic field vector. The GIST code \cite{xanthopoulos_geometry_2009} is used to calculate these flux-tube geometries. Due to the inclusion of $\omega_*^T$ in Eq.~(\ref{eq:aval}), $E_\mathrm{A}$ depends on the electron temperature and density gradients across the flux tube. These gradients are taken to be the typical experimentally observed quantities in QHS \cite{canik_experimental_2007, kumar_radial_2018}, defined as $a/L_{Te} = a/L_{ne} = 2$ for the electron temperature and density gradients, respectively. Note, $L_{\chi}$ is the characteristic gradient scale length defined as $L_{\chi} = -\chi/(d\chi/dr)$.
	
	The procedure for down-sampling the configurations then goes as follows. Using the three helical-shaping parameters, along with the quasi-helical symmetry metric, the database of configurations is partitioned into a set of selection bins that cover all metric values calculated from the equilibria. From each bin, a random set of configurations is selected in proportion to the number of configurations in that bin. This prevents over- or under-sampling any particular region of the configuration space. Importantly, configurations with rotational transforms that cross the 4/4 or 8/7 resonance are excluded, as such configurations can form large islands inside the plasma confinement region which contravene the VMEC solutions (see Sec.~5 in Ref.~\cite{Gerard_2023}).
	
\begin{figure}
		\centering
		\includegraphics[width=.48\textwidth, keepaspectratio]{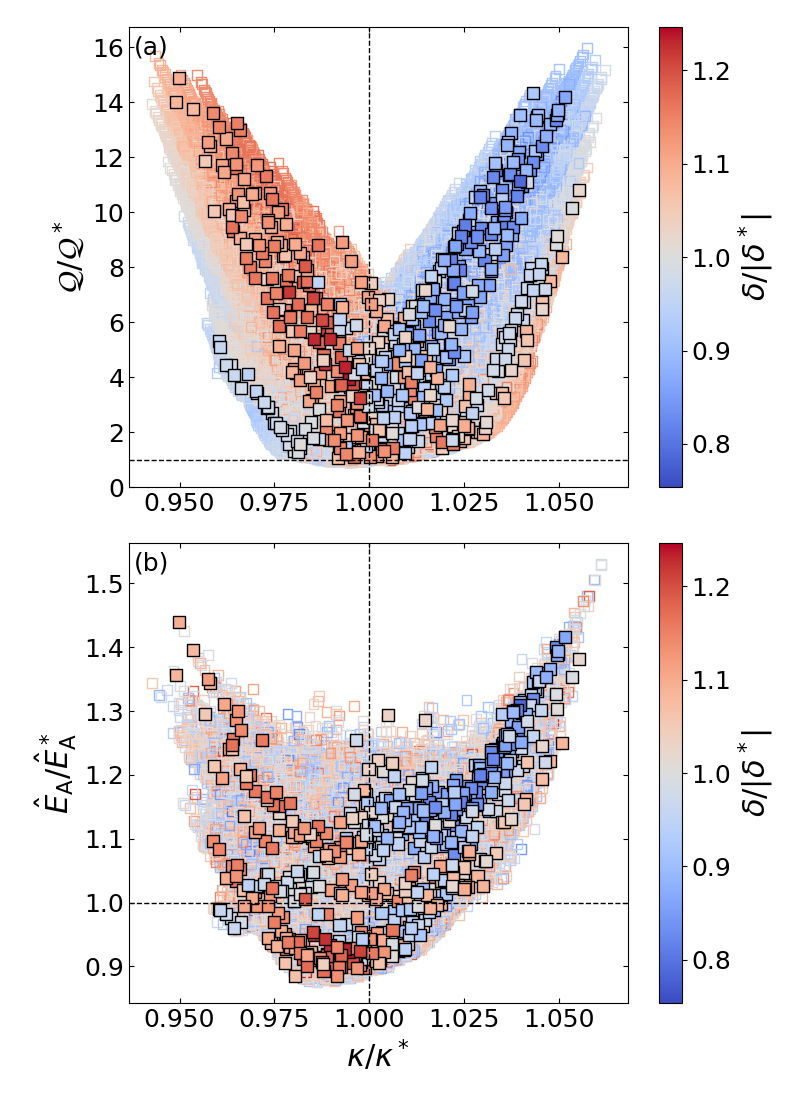}
		\caption{Symmetry-breaking and $E_\mathrm{A}$ ratios as functions of plasma elongation in panels (a) and (b), respectively, with triangularity shown as colors. The down-sampled set of 600 configurations is shown as filled squares in the foreground, while all other configurations in the database are shown as hollow squares in the background. Otherwise, the layering of configurations is random. The QHS configuration is indicated by the black dashed cross-hairs. \label{fig:down-sample}}
	\end{figure}

	Figure~\ref{fig:down-sample} shows the quasi-helical symmetry metric in (a) and the flux-tube $E_\mathrm{A}$ in (b), each as a function of elongation, with triangularity shown in color. The set of 600 down-sampled configurations is shown as filled squares in the foreground, while all other configurations in the database are shown as hollow squares in the background. Otherwise, the layering of configurations in each set is random. The black cross-hairs denote the metric values of the QHS flux surface in each figure. Note that thousands of configurations exist with comparable, or better, quasi-helical symmetry than QHS, and available energies are observed that exhibit a 10\% reduction from QHS. In Fig.~\ref{fig:down-sample}(a) the down-sampled configurations are seen to probe the full extent of the metric values, except for gaps found in regions with more extreme values of elongation as $\mathcal{Q}$ is increased. These gaps result from the exclusion of configurations with rotational transforms near the aforementioned resonances, meaning these excluded configurations are unlikely to make good experimental candidates. Moreover, in Fig.~\ref{fig:down-sample}(b) the down-sampled configurations are observed to span the range of $\hat{E}_\mathrm{A}/\hat{E}_\mathrm{A}^*$ from $0.89$ to $1.45$ in both $\kappa$ and $\delta$ shaping regimes. This demonstrates that the set of 600 configurations provides a representative sampling of the metric values produced in the database.
	
	\subsection{Equilibrium Reconstruction \label{sec:selection_results}}
	
	To calculate an ideal MHD equilibrium, VMEC requires an initial guess for the LCFS geometry and the enclosed toroidal magnetic flux $\psi_{\mathrm{edge}}$. When constructing the original database, $\psi_{\mathrm{edge}}$ was held fixed, and the LCFS of the QHS configuration was used as the initial guess. Then VMEC was run in free-boundary mode, meaning the initial guess for the LCFS was allowed to change based on how the coil-current information modified the vacuum magnetic field, with the LCFS expanding or contracting in volume to achieve the specified $\psi_{\mathrm{edge}}$. Moreover, solutions were provided using eight toroidal and eight poloidal mode numbers, allowing for computational expediency when generating the database.
	
	To ensure a robust geometric comparison in the proceeding gyrokinetic analysis, higher spectral resolutions are desired. However, VMEC often struggles to find free-boundary solutions in HSX geometries with poloidal or toroidal mode numbers greater than twelve. Alternatively, running VMEC in fixed-boundary mode allows for much higher mode resolution, but requires that a reliable LCFS be supplied as input since this boundary will be fixed as the solution is identified. Therefore, the LCFS of each down-sampled configuration is identified using a magnetic field-line-following (FLF) code in the corresponding vacuum-field configuration. This LCFS is then used to re-generate the VMEC equilibrium in fixed-boundary mode with 24 and 18 toroidal and poloidal mode numbers, respectively. The use of vacuum FLF is justified as the plasma pressure $p$ in HSX is low enough to warrant the $\beta = 0$ assumption, where $\beta = 2\mu_0p/B^2$ with $\mu_0$ the permeability of free space, making $\beta$ the ratio of thermal pressure to the magnetic-field pressure.
	
	Initially, the same $\psi_{\mathrm{edge}}$ is used in all high-resolution VMEC calculations, and the flux-surface-averaged magnetic field strength $\langle B \rangle_i$ is calculated on each $s_i$ surface. Then, VMEC is run a second time in fixed-boundary mode with the same mode resolution but with $\psi_{\mathrm{edge}, i} = (\langle B\rangle_\mathrm{q}/\langle B\rangle_i) \psi_\mathrm{edge,q}$. This results in a set of $s_i$ surfaces with the same flux-surface-averaged field strength. As a final check, the high-resolution $s_i$ surfaces are compared against the corresponding vacuum-field surface produced by FLF. This is to ensure the VMEC surfaces are not distorted by the presence of magnetic islands near those surfaces. In total, 563 configurations have been retained of the 600 configurations initially sampled. For reference, the high-resolution QHS metrics evaluate to $\kappa^* = 1.471$, $\delta^* = 0.0523$, $\zeta^* = -0.0901$, $\mathcal{Q}^* = 0.0089$, and $\hat{E}_\mathrm{A}^* = 0.0270$.
	
\begin{figure}
		\centering
		\includegraphics[width=.75\textwidth, keepaspectratio]{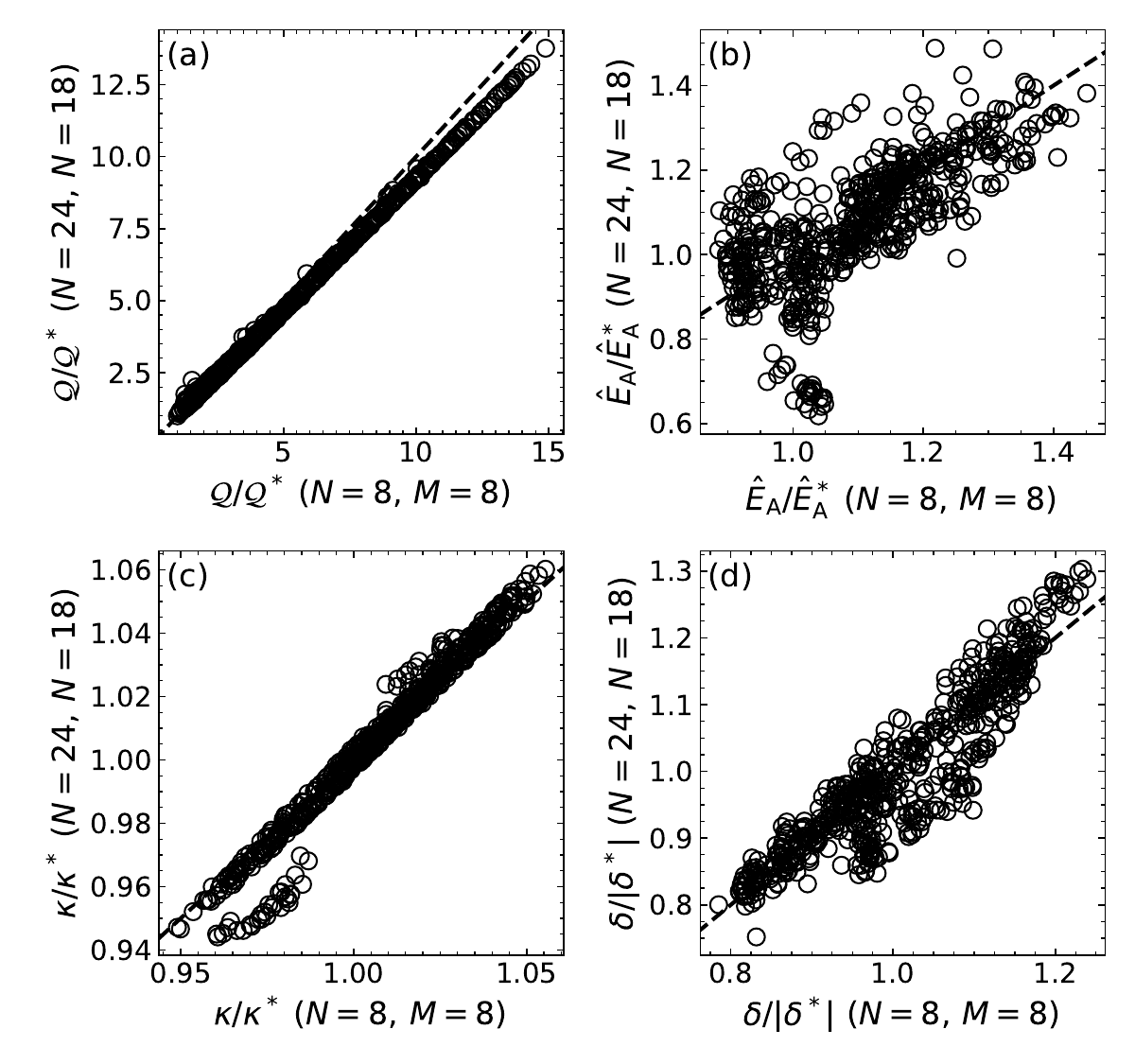}
		\caption{Metric quantities are compared in the fixed- and free-boundary solutions for the ideal MHD equilibria, where $(N=24,\,M=18)$ and $(N=8,\,M=8)$ resolutions are used, respectively. The metric quantities under consideration are the quasi-helical symmetry (a), flux-tube $E_\mathrm{A}$ (b), elongation (c), and triangularity (d). Note that the dashed black line indicates the hypothetical agreement between metric quantities calculated from both equilibrium solutions. \label{fig:compare}}
	\end{figure}

	With the metric quantities recalculated for the new equilibrium solutions, it is informative to compare them to their original values. Such a comparison is shown in Fig.~\ref{fig:compare}, where metric quantities are compared in the fixed- and free-boundary mode solutions. Note that the metric quantities calculated from the high-resolution fixed-boundary solutions are plotted along the vertical axis while the metrics calculated from the low-resolution free-boundary solutions are plotted along the horizontal axis. The dashed black line indicates perfect agreement between metric quantities calculated from both solutions. The metrics being considered are the quasi-helical symmetry metric (a), flux-tube $E_\mathrm{A}$ (b), elongation (c), and triangularity (d).
	
	It can be observed that both the quasi-helical symmetry and elongation values are nearly identical when calculated from either equilibrium solution. Alternatively, both the $E_\mathrm{A}$ and triangularity show considerably more scatter between their high- and low-resolution values. Note that squareness, not shown in the figure, is found to have no discernible trend between the high- and low-resolution values since the low-resolution equilibria fail to capture large values of squareness.
	
	Regarding the shaping parameters, these results demonstrate the intuitive notion that elongation is determined almost exclusively by low-order modes, while triangularity and squareness are related to progressively higher-order modes. Less intuitive, however, is that quasi-helical symmetry, like elongation, is determined by low-order modes, while $E_\mathrm{A}$ depends strongly on moderate- to high-order modes. This latter result is an important observation for the use of $E_\mathrm{A}$ as an optimization metric, showing that it requires high-resolution equilibria for an accurate calculation. This does not, however, cast doubt on the process for down-sampling the equilibria, since the high-resolution quantities are observed to span a range of metric values comparable to that found in the low-resolution equilibria. Therefore, the approach of investigating the equilibrium space around an existing configuration using the rapid generation of magnetic equilibria remains a valid optimization procedure. Lastly, the dependence of quasi-helical symmetry and $E_\mathrm{A}$ on low- and high-order modes, respectively, may provide some benefit regarding TEM optimization. For example, one could target quasi-helical symmetry by modifying the low-order modes and then perform a fine-tuning of the resultant configuration by targeting higher-order modes to enhance TEM stability. Additional support for this notion is presented towards the end of Sec.~\ref{sec:resonance}.
	
	\section{Linear Gyrokinetic Analysis \label{sec:gyrokinetic}}
	
	A set of linear gyrokinetic simulations across all 563 configurations and QHS are performed using the \textsc{Gene} code \cite{noauthor_gene_nodate, jenko_electron_2000}. The binormal wavenumbers investigated include $k_y\rho_\mathrm{s} =$~0.1, 0.4, 0.7, and 1. This choice of wavenumbers is motivated by nonlinear simulations in QHS, where it has been observed that the heat flux spectrum peaks at $k_y<1$ \cite{faber_gyrokinetic_2015}. The flux-tube geometries used in these simulations are selected from the $\alpha = 0$ field line and generated using the GIST code \cite{xanthopoulos_geometry_2009}. Convergence testing was done for 10 configurations selected from the set of 563, and growth rates $\gamma$ are converged to within $5\%$ deviations for a doubling of any numerical parameter or hyper-diffusion coefficient. The most stringent requirements for convergence are used in all simulations for a given $k_y$. For $k_y\rho_\mathrm{s} =$~0.7~and~1, the parameters are $N_x \times N_z \times N_{v\parallel} \times N_{\mu} = 15 \times 1024 \times 32 \times 8$, where $N_x$ and $N_z$ are the number of radial and parallel grid points, respectively, and $N_{v\parallel}$ and $N_{\mu}$ are the number of parallel and perpendicular velocity-space grid points, respectively. The velocity-space box sizes are $L_{v\parallel} = 3$ and $L_{\mu} = 9$ and the flux tube extends for 8 poloidal turns. The parallel and velocity-space hyper-diffusion coefficients, see Ref.~\cite{Pueschel_2010}, were set to $\epsilon_z=8$ and $\epsilon_v=0.2$, respectively. For $k_y\rho_\mathrm{s} = 0.4$ all parameters are the same except the flux tube extends for 16 poloidal turns and $N_z = 2048$. For $k_y\rho_\mathrm{s} = 0.1$, the parameters are identical to the $k_y\rho_\mathrm{s} = 0.4$ case except that $N_{v\parallel} = 64$. Note that $E_\mathrm{A}$ presented in this section have been calculated in the same poloidally extended flux-tubes as are used for the \textsc{Gene} simulations, though no significant differences were observed relative to $E_\mathrm{A}$ calculations performed in the four-poloidal-turn flux-tubes. Physical parameters in the simulations are $\beta = 0$, $\nu_\mathrm{ei} = 0$, $T_\mathrm{i}/T_\mathrm{e} = 0.2$, $a/L_\mathrm{ne} = a/L_\mathrm{ni} = a/L_\mathrm{Te} = 2$ and $a/L_\mathrm{Ti} = 0$, where $\nu_\mathrm{ei}$ is the electron-ion collision frequency. This set of parameters is chosen to reflect the physical parameters observed in typical experiments \cite{canik_experimental_2007, kumar_radial_2018}.
	
\begin{figure}
		\centering
		\includegraphics[width=.75\textwidth, keepaspectratio]{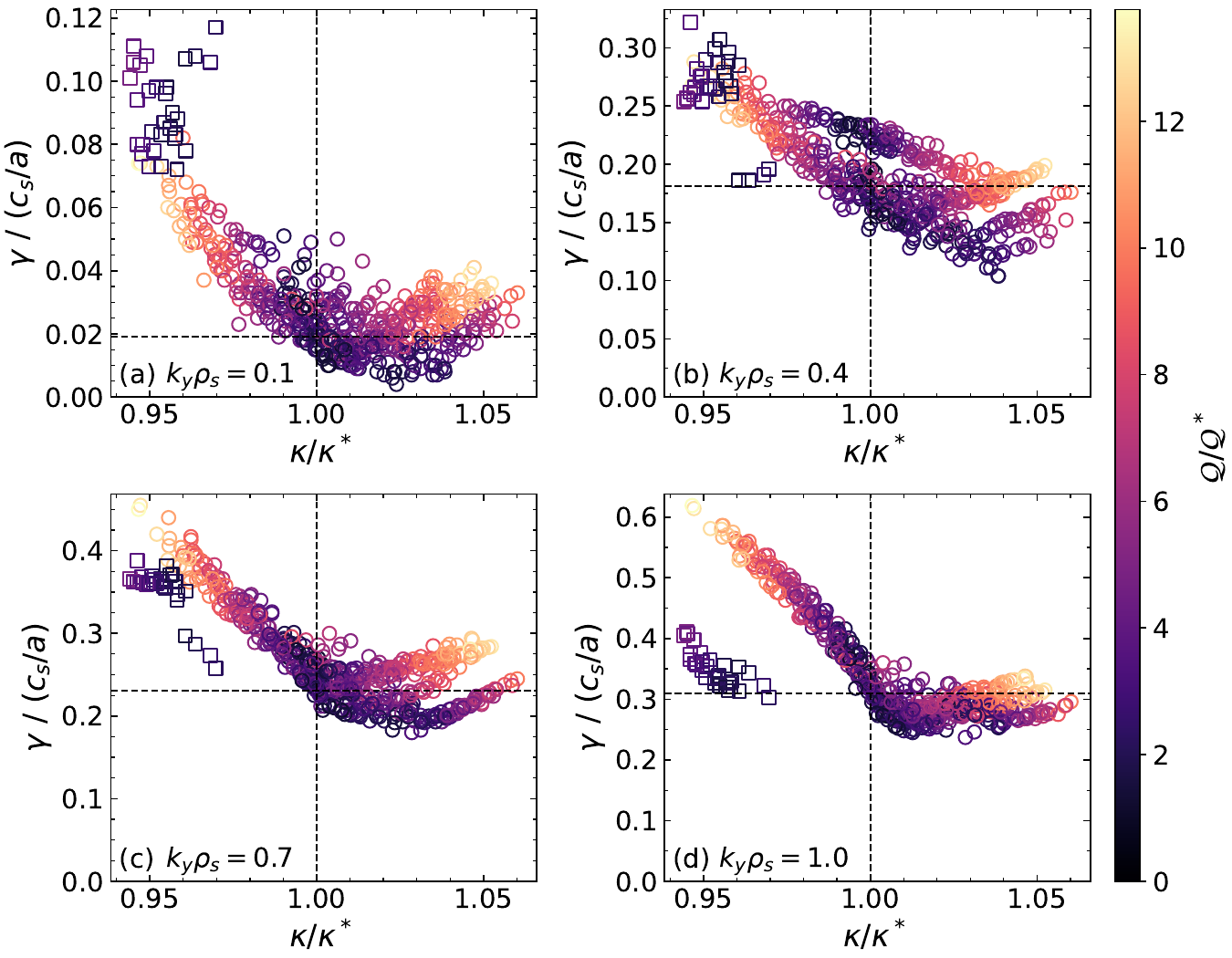}
		\caption{Growth rates are shown as a function of plasma elongation at binormal wavenumbers 0.1, 0.4, 0.7, and 1 in panels (a), (b), (c), and (d), respectively, with quasi-helical symmetry ratios shown in color. The black cross-hairs indicate the QHS configuration in each panel and the square markers are a set of large-negative-squareness configurations. The lowest growth rates are typically found in configurations with high elongation and good quasi-helical symmetry. \label{fig:gene_vs_kappa}}
	\end{figure}

	Figure~\ref{fig:gene_vs_kappa} shows the TEM growth rates $\gamma$ of the most unstable mode as a function of plasma elongation, with $\gamma$ normalized to the ion sound speed $c_\mathrm{s}$ divided by effective minor radius, and quasi-helical-symmetry ratios shown in color. Panels (a), (b), (c), and (d) correspond to wavenumbers $k_y\rho_\mathrm{s} =$~0.1,~0.4,~0.7,~and~1, respectively. The location of the QHS configuration is identified in each panel by the black dashed cross-hairs. It can be observed that the lowest growth rates, across all wavenumbers, tend to occur in configurations with high elongation and low symmetry-breaking amplitudes. Relative to QHS, the minimal growth rates are reduced by a factor of 4.8 and 1.7 at high elongation for $k_y\rho_\mathrm{s} = 0.1$ and 0.4, respectively. Relative to the largest observed growth rates, the reduction factor is 29 and 3.1 for $k_y\rho_\mathrm{s} = 0.1$ and 0.4, respectively. These results are consistent with the gyrokinetic results from Ref.~\cite{Gerard_2023} regarding the stabilizing effect of plasma elongation. However, these new results further demonstrate that the reduction in $\gamma$ with increasing $\kappa$ is limited by the ability to increase $\kappa$ while preserving the quasi-helical symmetry. This is exemplified by the increase in growth rates at high elongation and increasingly broken quasi-helical symmetry.
	
\begin{figure}
		\centering
		\includegraphics[width=.48\textwidth, keepaspectratio]{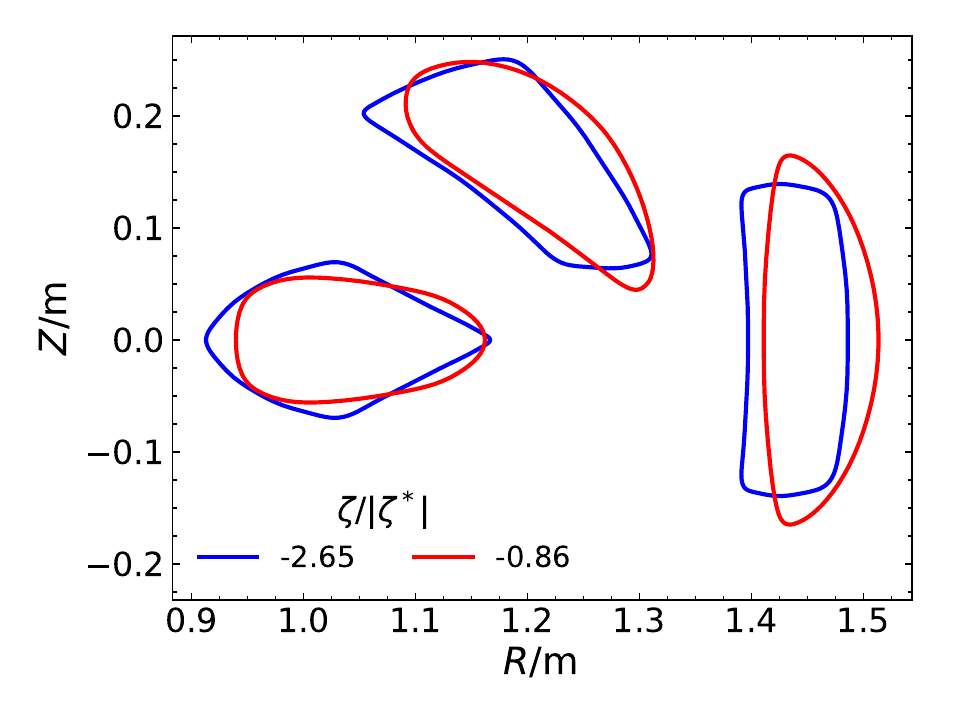}
		\caption{Three toroidal cross-sections of two exemplary flux surfaces. This set of surfaces represents the extrema in squareness of the down-sampled database, with the maximum and minimum squareness shown in red and blue, respectively. \label{fig:squareness}}
	\end{figure}

	In Fig.~\ref{fig:gene_vs_kappa}, a set of outlier configurations is present at all wavenumbers and is distinguished from the bulk configurations with square markers. Notably, in Fig.~\ref{fig:gene_vs_kappa}(a) and (b) these configurations are observed to have the largest growth rates of all configurations, while in Fig.~\ref{fig:gene_vs_kappa}(c) and (d), their growth rates are reduced relative to other configurations with comparable elongation. Also, when compared against the bulk configurations, they demonstrate anomalously good quasi-helical symmetry for their low elongation. Interestingly, these configurations are observed to have the most negative squareness values of all 564 configurations considered, with ratios $\zeta/|\zeta^*| \leq -1.5$. These large-absolute-value squareness configurations are the result of a $4/4$ resonance near the plasma edge, generating diamond-shaped flux surfaces in the helically rotated reference frame. This is exemplified in Fig.~\ref{fig:squareness}, where three toroidal cross-sections are shown for two different flux surfaces. These flux surfaces represent the extrema in squareness in the down-sampled database, with the maximum and minimum surfaces shown in red and blue, respectively. The outlier nature of the large-negative-squareness configurations will be observed throughout this section, and are represented with square markers throughout the paper.
	
\begin{figure}
		\centering
		\includegraphics[width=.75\textwidth, keepaspectratio]{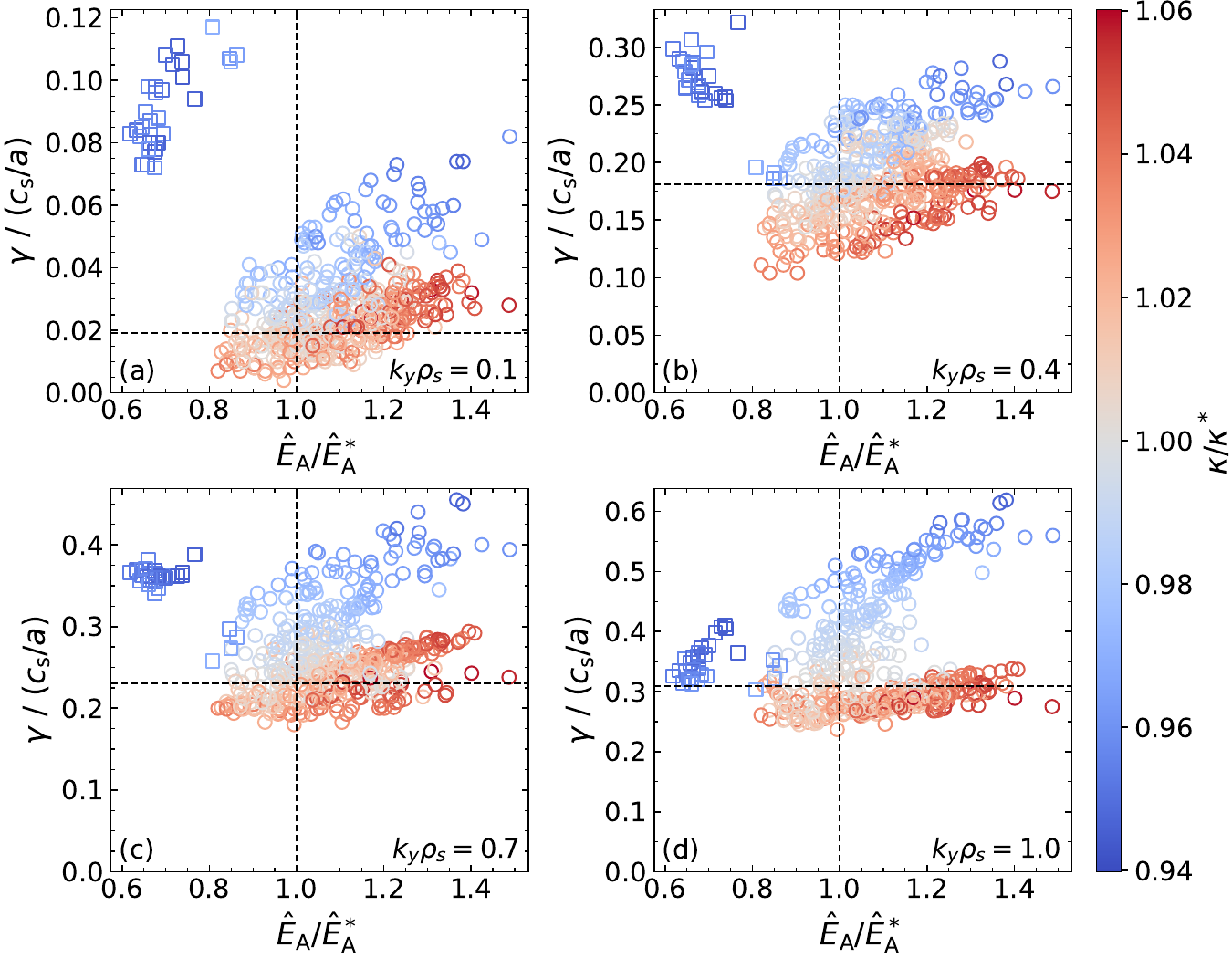}
		\caption{Growth rates are shown as a function of the flux-tube $E_\mathrm{A}$ at binormal wavenumbers 0.1, 0.4, 0.7, and 1 in panels (a), (b), (c), and (d), respectively, with elongation ratios shown in color. The black cross-hairs indicate the QHS configuration in each panel and the square markers are a set of large-negative-squareness configurations. A robust scaling between growth rates and $E_\mathrm{A}$ is only observed when the different elongation branches are taken into consideration separately. \label{fig:gene_vs_ae}}
	\end{figure}

	To investigate how growth rates compare to $E_\mathrm{A}$, Fig.~\ref{fig:gene_vs_ae} shows the growth rates as a function of the flux-tube $E_\mathrm{A}$, with plasma elongation shown in color. Configurations with elongation greater than in QHS are plotted in shades of red, and lower-elongation configurations are plotted in blue. Binormal wavenumbers 0.1, 0.4, 0.7, and 1 are shown in panels (a), (b), (c), and (d), respectively, with the QHS configuration indicated with the black cross-hairs.
	
	Without considering plasma elongation, a scattered correlation between growth rates and $E_\mathrm{A}$ can be observed, exemplified by the wide range of growth rates at any given value of $E_\mathrm{A}$. However, if elongation is considered, one observes two separate branches of configurations, with each branch delineated by elongations higher or lower than that observed in the QHS flux surface. These separate branches exhibit a less scattered trend, showing reduced growth rates for lower $E_\mathrm{A}$. This suggests that, as a linear metric, $E_\mathrm{A}$ may be limited in its ability to identify a globally optimized configuration, but that within a particular shaping regime, it may be used as a local optimization metric. For example, if one were to begin optimization with an initial configuration appearing in the top-right quadrant of Fig.~\ref{fig:gene_vs_ae}(b), and then smoothly deform the flux surface to reduce the $E_\mathrm{A}$, one would likely arrive at a configuration that has a lower growth rate. However, there could still be a considerable disparity between that growth rate and another growth rate from a configuration with a very similar $E_\mathrm{A}$. Furthermore, the set of outlier configurations with large squareness is identified by their exceptionally low $E_\mathrm{A}$ and their uncharacteristically large growth rates at some $k_y$. The reason for the anomalously large growth rates in relation to the $E_\mathrm{A}$ is unknown. However, it is possible that these configurations constitute a third shaping regime, distinct from the high- and low-elongation regimes, in which separate scaling relations could exist between $\gamma$ and $E_\mathrm{A}$.
	
\begin{figure}
		\centering
		\includegraphics[width=.75\textwidth, keepaspectratio]{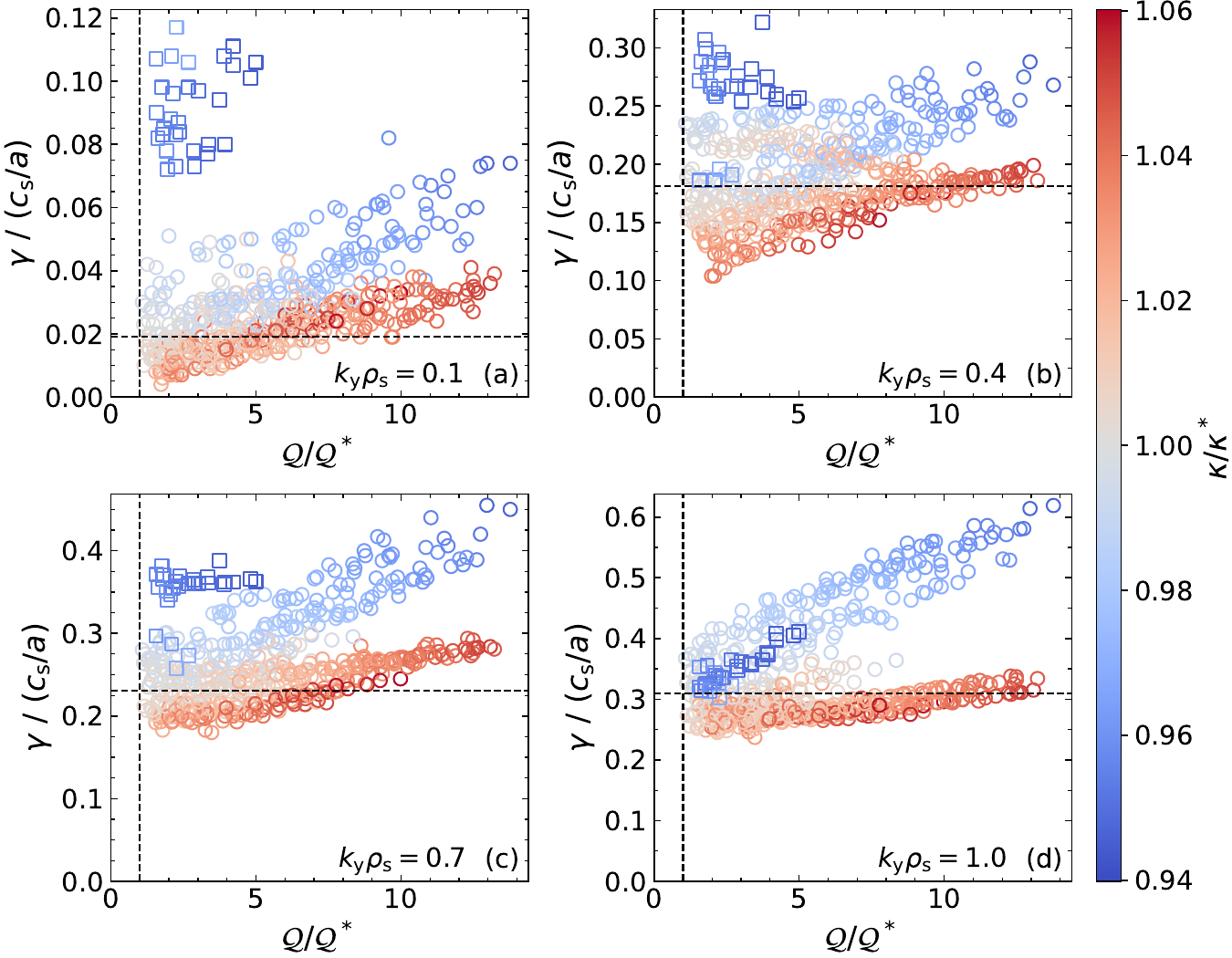}
		\caption{Growth rates are shown as a function of symmetry breaking ratios at binormal wavenumbers 0.1, 0.4, 0.7, and 1 in panels (a), (b), (c), and (d), respectively, with elongation ratios shown in color. The black cross-hairs indicate the QHS configuration in each panel and the square markers are a set of large-negative-squareness configurations. Similar trends as observed in Fig.~\ref{fig:gene_vs_ae} are observed here, though the trends with respect to their quasi-helical symmetry are less scattered. \label{fig:gene_vs_qhs}}
	\end{figure}

	To relate the $E_\mathrm{A}$ of these configurations to their quasi-helical symmetry, Fig.~\ref{fig:gene_vs_qhs} shows the growth rates as a function of the symmetry-breaking ratio, with elongation shown in color. Similar to before, wavenumbers 0.1, 0.4, 0.7, and 1 are shown in panels (a), (b), (c), and (d), respectively, with the QHS configuration identified by the black cross-hairs. By comparing Figs.~\ref{fig:gene_vs_ae} and \ref{fig:gene_vs_qhs}, similar trends can be identified; however, the scaling between $\gamma$ and $\mathcal{Q}$ in Fig.~\ref{fig:gene_vs_qhs} appears less scattered. A certain degree of consistency between $E_\mathrm{A}$ and quasi-helical symmetry is to be expected, given that minimizing $\overline{\omega}_{\psi}$ is one method by which $E_\mathrm{A}$ can be reduced, as seen in Eq.~(\ref{eq:aval}). What is more surprising is that in the comparison of Figs.~\ref{fig:gene_vs_ae} and \ref{fig:gene_vs_qhs}, the quasi-helical symmetry appears to be a better predictor of the relative growth rate than $E_\mathrm{A}$. Of course, the set of outlier configurations is still present, appearing in Fig.~\ref{fig:gene_vs_qhs} as the set of configurations with the lowest elongations and $\mathcal{Q}/\mathcal{Q}^* \leq 5$.
	
	\section{TEM Resonance Analysis \label{sec:resonance}}
	
	\subsection{Deriving the velocity-space-averaged TEM resonance operator}
	To better understand these results, and to compare them to existing analytic theory, a velocity-space-averaged resonance operator is defined and calculated from the linear gyrokinetic data. This resonance operator is motivated by Eq.~(\ref{eq:power_transfer}), restated here for convenience
	\begin{equation*}
		P = \frac{\pi e^2}{T_0} \int_{-\infty}^{\infty} \frac{d\ell}{B} \int_{-\infty}^{\infty} \omega \left( \omega - \omega_*^T \right) \delta\left( \omega - \overline{\omega}_\mathrm{d} \right) |\overline{\phi}|^2 f_0 d^3v.
	\end{equation*}
	Note that $\omega(\omega-\omega_*^T)$ is the salient quantity regarding the influence of trapped-electron drifts on the TEM stability of a particular flux tube. Therefore, the velocity-space-averaged resonance operator
	\begin{equation}
		\langle \omega(\omega-\omega_*^T) \rangle = \frac{ \int_{-\infty}^{\infty} \omega(\omega-\omega_*^T) \delta(\omega - \overline{\omega}_\mathrm{d}) |\overline{\phi}|^2 f_0 d^3v }{ \int_{-\infty}^{\infty} \delta(\omega - \overline{\omega}_\mathrm{d}) |\overline{\phi}|^2 f_0 d^3v } \label{eq:velocity_average}
	\end{equation}
	is defined. By evaluating this quantity, one may characterize the contribution of the trapped-electron drifts to the destabilizing transfer of energy described by $P < 0$. The evaluation will be carried out by first making a series of substitutions to express the integral in terms of particle energy and pitch angle, and then normalizing the equation variables to be consistent with \textsc{Gene}. The evaluation of the Dirac-delta distribution function will then be made possible with the drift-wave frequency $\omega$ calculated in \textsc{Gene}, which will result in the reduction of Eq.~(\ref{eq:velocity_average}) to a field-line integral.
	
	The velocity differential can be transformed to a particle energy and pitch angle differential as described in Ref.~\cite{roach_trapped_1995} as
	\begin{equation}
		d^3v = \pi \left( \frac{2T_0}{m} \right)^{3/2} \varepsilon^{1/2} p(\theta_\mathrm{b}) d\varepsilon d\theta_\mathrm{b}, \label{eq:vel_diff}
	\end{equation}
	where $\varepsilon = E/T_0$ is the normalized particle energy, $\theta_\mathrm{b}$ is the poloidal angle where a trapped particle is bounced -- therefore relating it to the pitch angle -- and the phase-space-density factor is
	\begin{equation}
		p(\theta_\mathrm{b}) = \frac{|\mathcal{B}^{\prime}(\theta_\mathrm{b})|}{\sqrt{1-\mathcal{B}(\theta_\mathrm{b})}},
	\end{equation}
	with $\mathcal{B} = B_{\mathrm{min}}/B(\theta_\mathrm{b})$, $B_{\mathrm{min}}$ the minimum magnetic-field strength over the poloidal domain, and $\mathcal{B}^{\prime} = d\mathcal{B}/d\theta$. Importantly, $p(\theta_\mathrm{b})$ is proportional to the trapped-particle density, and peaks when $B = B_\mathrm{min}$. Then, a Maxwellian is used for the zeroth-order distribution function,
	\begin{equation}
		f_0 = n_0 \left( \frac{m}{2\pi T_0} \right)^{3/2} e^{-\varepsilon}. \label{eq:maxwellian}
	\end{equation}
	
	To express Eq.~(\ref{eq:velocity_average}) in a manner consistent with \textsc{Gene}, first consider that the drift-wave frequency and wavenumber normalizations are
	\begin{align}
		\omega &= \left( \frac{c_\mathrm{s}}{a} \right) \hat{\omega} &
		k_{\psi} &= \left( \frac{q_0}{2 \psi_\mathrm{edge} \rho_*     \sqrt{s_0} } \right) \hat{k}_\mathrm{x} &
		k_{\alpha} &= \left( \frac{\sqrt{s_0}}{q_0 \rho_*} \right)     \hat{k}_\mathrm{y}, \label{eq:gene_normalizations}
	\end{align}
	where the $0$ subscript denotes the variable is evaluated along the field line at the center of the flux tube, $q_0$ is the safety factor, and $\hat{\omega}$, $\hat{k}_\mathrm{x}$, and $\hat{k}_\mathrm{y}$ are the corresponding dimensionless values in \textsc{Gene} \cite{beer_1995, noauthor_gene_nodate}. Then, the diamagnetic-drift frequency is
	\begin{equation}
		\omega_*^T = -\left( \frac{T_0}{2e     q_0\psi_\mathrm{edge} \rho_*} \right) \hat{\omega}_*^T, \label{eq:norm_omg_str}
	\end{equation}
	with $\hat{\omega}_*^T = \hat{k}_\mathrm{y}[(a/L_{ne}) + (a/L_{Te})(z-3/2)]$, and the bounce-averaged-drift frequencies are
	\begin{align}
		\overline{\omega}_{\alpha} &= \left(     \frac{T_0}{2e q_0\psi_\mathrm{edge}\rho_*} \right) \hat{\overline{\omega}}_\mathrm{y} \label{eq:norm_omg_alf} \\
		\overline{\omega}_{\psi} &= \left( \frac{q_0     T_0}{2e\psi_\mathrm{edge}\rho_*} \right) \hat{\overline{\omega}}_\mathrm{x}. \label{eq:norm_omg_psi}
	\end{align}
	Note that $\hat{\overline{\omega}}_\mathrm{y} = \varepsilon \hat{k}_\mathrm{y} \hat{\overline{v}}_\mathrm{dy}$ and $\hat{\overline{\omega}}_\mathrm{x} = \varepsilon\hat{k}_\mathrm{x} \hat{\overline{v}}_\mathrm{dx}$, with normalized drift velocities $\hat{v}_\mathrm{dy}$ and $\hat{v}_\mathrm{dx}$ defined in Eq.~(\ref{eq:alpha_drift_gist_final})--(\ref{eq:psi_drift_gist_final}). A general discussion on how these quantities are calculated from a GIST output is presented in Appendix~\ref{appdx:drifts}. Finally, the full bounce-average-drift frequency is
	\begin{equation}
		\overline{\omega}_\mathrm{d} = \left( \frac{T_0}{     2eq_0\psi_\mathrm{edge}\rho_* } \right) \left[ 1 + q_0^2 \left( \frac{ \hat{\overline{\omega}}_\mathrm{x} }{ \hat{\overline{\omega}}_\mathrm{y} } \right) \right] \hat{\overline{\omega}}_\mathrm{y}.
	\end{equation}
	
	Thus defined, the root of the Dirac-delta distribution in Eq.~(\ref{eq:velocity_average}) can be identified as
	\begin{equation}
		\varepsilon^{\prime}(\theta_\mathrm{b}) = \left( \frac{c_\mathrm{s}\rho_* }{ \xi a } \right) \left[ 1 + q_0^2 \left( \frac{ \hat{\overline{\omega}}_\mathrm{x} }{ \hat{\overline{\omega}}_\mathrm{y} } \right) \right]^{-1} \left( \frac{ \hat{\omega} }{ \hat{\overline{\omega}}_\mathrm{y} } \right),
	\end{equation}
	with $\xi = T_0/(2eq_0\psi_\mathrm{edge})$. Then, substituting Eq.~(\ref{eq:vel_diff})--(\ref{eq:maxwellian}) into Eq.~(\ref{eq:velocity_average}) and applying the \textsc{Gene} normalizations, Eq.~(\ref{eq:velocity_average}) is reduced to the field-line integral
	\begin{equation}
		\langle \omega(\omega-\omega_*^T) \rangle = \hat{\omega}^2 \left(\frac{c_\mathrm{s}}{a}\right)^2 \frac{ \int_{\theta_i}^{\theta_j} R(\theta_\mathrm{b}) H(\varepsilon^{\prime}) \sqrt{\varepsilon^{\prime}} \mathrm{e}^{-\varepsilon^{\prime}} |\overline{\phi}|^2 p(\theta_\mathrm{b}) d\theta_\mathrm{b} }{ \int_{\theta_i}^{\theta_j} \sqrt{\varepsilon^{\prime}} \mathrm{e}^{-\varepsilon^{\prime}} |\overline{\phi}|^2 p(\theta_\mathrm{b}) d\theta_\mathrm{b} }, \label{eq:average_reduced}
	\end{equation}
	with $H(\varepsilon^{\prime})$ the Heaviside function and
	\begin{equation}
		R(\theta_\mathrm{b}) = 1 + \left( \frac{\xi a}{c_\mathrm{s}\rho_*} \right) \left( \frac{\hat{k}_\mathrm{y}}{\hat{\omega}} \right) \left[ \frac{a}{L_\mathrm{n}} + \frac{a}{L_\mathrm{Te}} \left( \varepsilon^{\prime} - \frac{3}{2}\right) \right]. \label{eq:resonance_function}
	\end{equation}
	
\begin{figure}
		\centering
		\includegraphics[width=\textwidth, keepaspectratio]{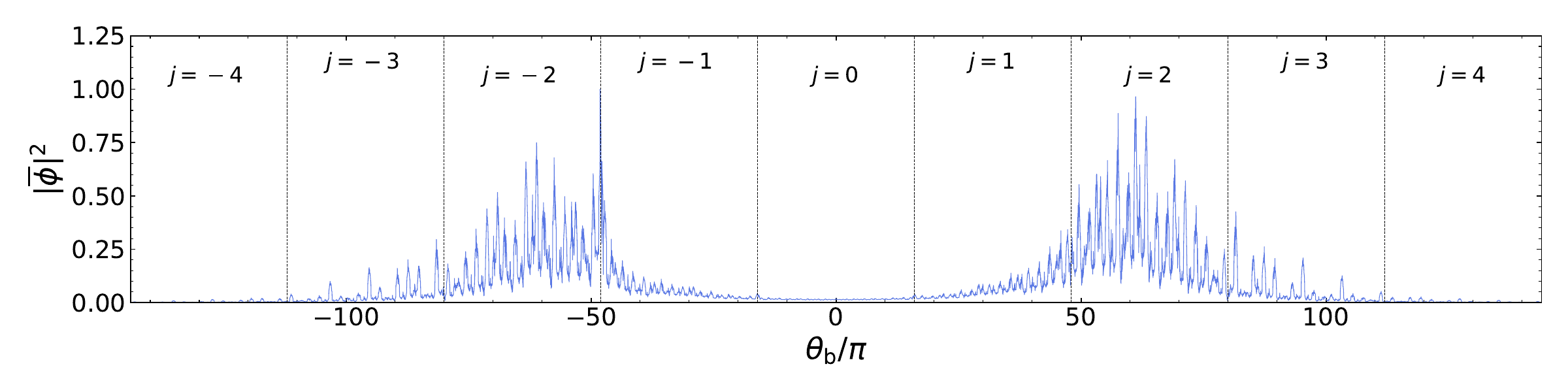}
		\caption{Bounce-averaged-electrostatic eigenfunction of the QHS configuration at $\hat{k}_\mathrm{y}=0.1$ as a function of poloidal-bounce angle in ballooning space. Each $32\pi$ poloidal extension of the magnetic field line about the central $j=0$ section is delineated with a vertical dashed line. Sections with a finite radial wavenumber are identified by $j\neq0$. \label{fig:eigenfucntion}}
	\end{figure}

	Importantly, radial wavenumber resonances in \textsc{Gene} are related to the ballooning representation of the electrostatic eigenfunction \cite{beer_1995}. This relation is shown graphically in Fig.~\ref{fig:eigenfucntion}, where the square of the bounce-averaged-eigenfunction amplitude in the QHS configuration at $\hat{k}_\mathrm{y} = 0.1$ is shown as a function of the poloidal-bounce angle in ballooning space. This choice of wavenumber is selected because in the derivation of Eq.~(\ref{eq:power_transfer}) it is assumed that $\gamma \rightarrow 0^+$, which is best fulfilled at low $k_\mathrm{y}$. Therefore, in the subsequent analysis, the focus will be on the $\hat{k}_\mathrm{y}=0.1$ simulations. The vertical bars in the figure indicate each $32\pi$ poloidal extension of the flux-tube domain about the central $j=0$ segment. These extensions are indexed with $j$, so that $\hat{k}_\mathrm{x} = (2j\pi)/L_\mathrm{x}$, with $L_\mathrm{x}$ the radial-box size of the flux tube in units of $\rho_\mathrm{s}$. In the figure, the eigenfunction is shown over domains with $j \in [-4, 4]$, though in simulation the domain extends to $j \in [-7, 7]$.
	
	From Fig.~\ref{fig:eigenfucntion}, a significant eigenfunction amplitude at finite-radial wavenumber can be observed, with the peak found at $|j| = 2$. This means the radial drifts cannot be neglected in Eq.~(\ref{eq:average_reduced}) a priori, especially when considering configurations with broken quasi-helical symmetry. However, as will be shown, the TEM resonance is predominantly determined by the precessional, rather than radial, drifts. Therefore, to understand how the precessional drifts influence the TEM resonance, it is helpful to define the velocity-space-averaged bounce-average precessional drift as
	
	\begin{equation}
		\{\overline{\omega}_{\alpha}\} = \frac{\int_{-\infty}^{\infty} \overline{\omega}_{\alpha} f_\mathrm{0} d^3v }{ \int_{-\infty}^{\infty} f_\mathrm{0} d^3v } = \frac{3}{2} \left( \frac{\xi\hat{k}_\mathrm{y}}{\rho_*} \right) \frac{\int_{\theta_i}^{\theta_j} \hat{\overline{v}}_\mathrm{dy} p(\theta_\mathrm{b}) d\theta_\mathrm{b} }{ \int_{\theta_i}^{\theta_j} p(\theta_\mathrm{b} d\theta_\mathrm{b} } \label{eq:alpha_drift_average}
	\end{equation}
	Here, Eqs.~(\ref{eq:vel_diff}), (\ref{eq:maxwellian}), and (\ref{eq:norm_omg_alf}) are used to reduce the velocity-space average to a field-line integral covering the poloidal domain.
	
	\subsection{TEM and UI cross-phase and resonance analysis}
	
	It is additionally illuminating to consider the cross-phase between the electrostatic potential and density fluctuations, with the trapped- and passing-electron populations considered separately. The reason for this separation is that TEM and UI cross phases are typically in proximity to $\pi/2$ in the trapped- and passing-electron populations, respectively. Therefore, by comparing the cross-phase data between modes, one may identify mode branches that have a TEM, UI, or mixed cross-phase. Cross phases are commonly calculated for each population of electrons as a histogram throughout all of phase space. Therefore, a cross-phase distribution must be considered. The process by which these histograms are reduced to a scalar is described in Appendix~\ref{appdx:cross_phase}; however, the salient quantity presented here is $M_{\phi\times n}$, which is shifted towards $-1$ when the cross phase has a strong UI signature, $+1$ when indicating a TEM, and $\approx 0$ when the mode is a mixed UI/TEM.    
	
\begin{figure}
		\centering
		\includegraphics[width=.98\textwidth, keepaspectratio]{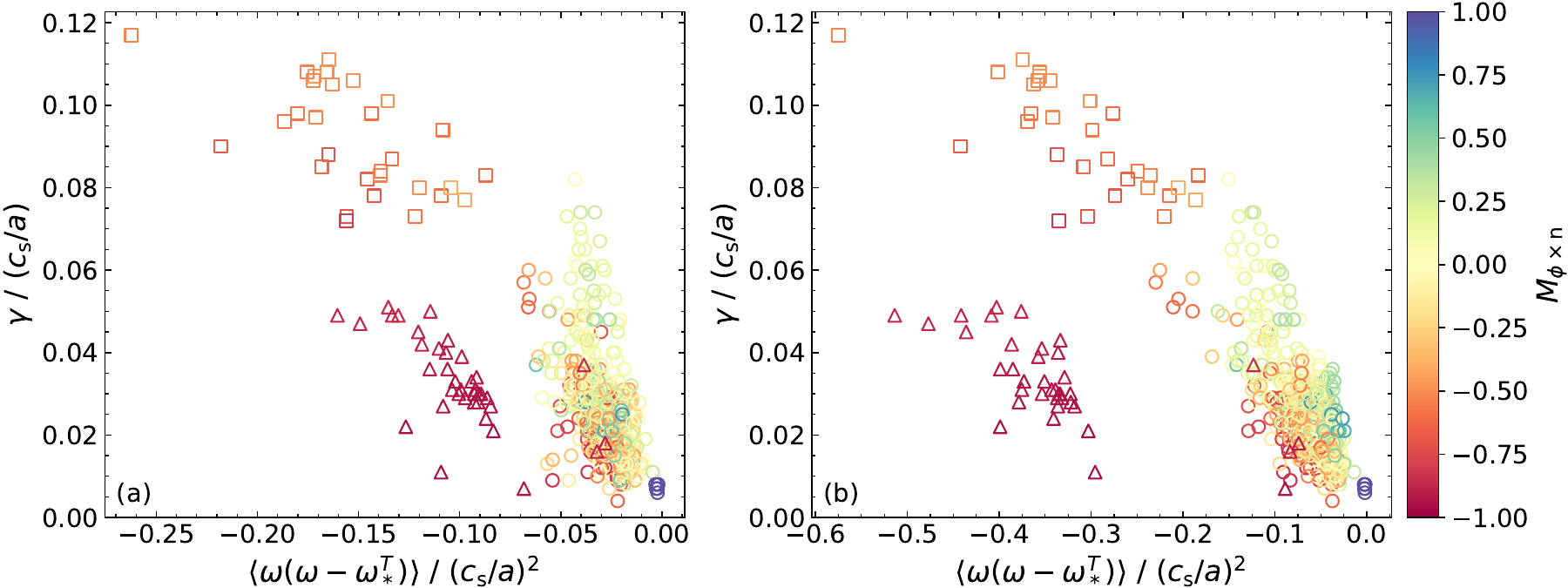}
		\caption{$k_y\rho_\mathrm{s} = 0.1$ growth rates are plotted as a function of the velocity-space-averaged TEM resonance operator, with the resonance calculated for all radial wavenumbers in panel (a) and only for $k_\mathrm{x}=0$ in panel (b). The color scale distinguishes TEMs from UI going from blue to red, respectively. The square markers denote the set of large-negative-squareness configurations and the triangle markers are a set of configurations with a strong UI cross phase. It is observed that the growth-rate reduction is consistent with analytic theory and that the resonance is determined primarily by the precessional, rather than radial, drifts. \label{fig:resonance}}
	\end{figure}

	Figure~\ref{fig:resonance} shows the maximal growth rates at $\hat{k}_\mathrm{y} = 0.1$ for all 564 configurations as a function of $\langle \omega (\omega - \omega_*^T) \rangle$, with $M_{\phi\times n}$ shown in color. In panel (a), all radial wavenumbers are included in the evaluation of Eq.~(\ref{eq:average_reduced}), while in panel (b), only the $\hat{k}_\mathrm{x} = 0$ mode is considered, thereby excluding information regarding the radial drifts. It can be observed that the reduction in growth rates, in both panels, is correlated with a shift of $\langle\omega(\omega-\omega_*^T)\rangle$ towards zero, demonstrating that the stabilization is consistent with the analytic theory of collisionless TEMs \cite{proll_resilience_2012, helander_collisionless_2013}.
	
	A set of outliers are presented with triangle markers in Fig.~\ref{fig:resonance}(b), and identified by their large negative values of $\langle \omega (\omega - \omega_*^T) \rangle$ and relatively low-growth rates. Due to the strong UI signature of these modes, with $M_{\phi\times n} < -0.86$, one does not expect the velocity-space-averaged TEM resonance operator to accurately capture their stabilization. However, a similar trend is observed with these outliers as is observed in the bulk configurations, namely that the growth rates are reduced as the resonance shifts towards zero. Interestingly, the outlying nature of these configurations is minimized when all finite radial wavenumbers are included, as seen in Fig.~\ref{fig:resonance}(a). Moreover, the cross-phase metric for these configurations is shifted towards $-1$ as the TEM resonance is reduced, with the lowest outlier growth rate exhibiting $M_{\phi\times n}=-0.94$. This shows these modes are increasingly UI dominant as the TEM resonance is reduced. Alternatively, no clear transition from a TEM to a UI cross-phase is observed in the bulk configurations. To be sure, if one disregards the outlier configurations, the strongest UI signatures are found near these zero crossings, but so too are the strongest TEM signatures. A likely explanation for this is that the UI is stronger in the outlier configurations, making the transition to UI dominance more accessible with a reduction in the TEM resonance. Interestingly, these outliers do not exhibit any unique properties regarding their helical shaping parameters, meaning they are decidedly not the same set of outliers identified in Sec.~\ref{sec:gyrokinetic}. This means the resonance operator accurately captures the resonance properties of the large-negative-squareness configurations, which are identified by their square markers. This further highlights the inability of $E_\mathrm{A}$ to characterize TEM growth rates in various shaping regimes.
	
	In Ref.~\cite{costello_2023}, the configuration that exhibited the clearest transition from a TEM to UI cross phase was the W7-X configuration most congruous with the maximum-$\mathcal{J}$ condition. Therefore, to reconcile the observed reduction in the TEM resonance operator without this corresponding transition, it is informative to consider how well these configurations satisfy $\partial \mathcal{J} / \partial \psi < 0$. From Eq.~(\ref{eq:precessional_drift}) it is known that the sign of $\overline{\omega}_{\alpha}$ is determined exclusively from $\partial \mathcal{J}/\partial \psi$. Therefore, Eq.~(\ref{eq:alpha_drift_average}) can be used as a proxy for maximum-$\mathcal{J}$-ness.
	
\begin{figure}
		\centering
		\includegraphics[width=.48\textwidth, keepaspectratio]{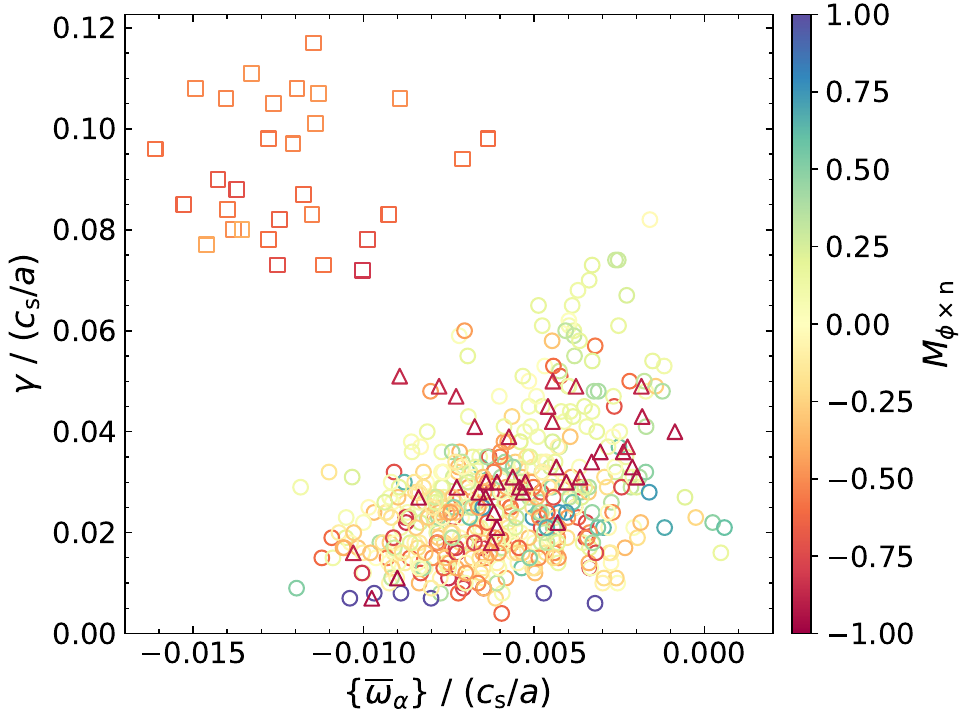}
		\caption{$k_y\rho_\mathrm{s} = 0.1$ growth rates are plotted as a function of $\{\overline{\omega}_{\alpha}\}$, while the color scale distinguishes TEMs from UI going from blue to red, respectively. This shows that no configuration is likely to satisfy $\partial \mathcal{J} / \partial \psi < 0$, which suggests the configurations may not be sufficiently TEM-stabilized to observe the transition from a dominant TEM to UI. \label{fig:gene_drifts}}
	\end{figure}

	Figure~\ref{fig:gene_drifts} shows the $\hat{k}_\mathrm{y}=0.1$ growth rates as a function of $\{\overline{\omega}_{\alpha}\}$, with the cross-phase signature $M_{\phi\times n}$ shown in color. It is of significant note that no clear trend appears in the comparison between the growth rates and velocity-space-averaged precessional drift or the cross-phase signature. Moreover, the sign of $\{\overline{\omega}_{\alpha}\}$ indicates that $\partial \mathcal{J}/\partial \psi >0$ is likely for all configurations, meaning the maximum-$\mathcal{J}$ condition is never observed. The implications here are twofold. First, the reduction in the TEM resonance results from the detailed interaction between the trapped-particle fraction, electrostatic eigenfunction, and the bounce-averaged precessional drifts rather than a general trend towards drift reversal along the entire flux tube. Second, the drifts that persist in the destabilizing electron-diamagnetic direction are sufficient to prevent the dominance of the UI over the TEM.
	
	\subsection{Impact of elongation on TEM resonance}
\begin{figure}
		\centering
		\includegraphics[width=.75\textwidth, keepaspectratio]{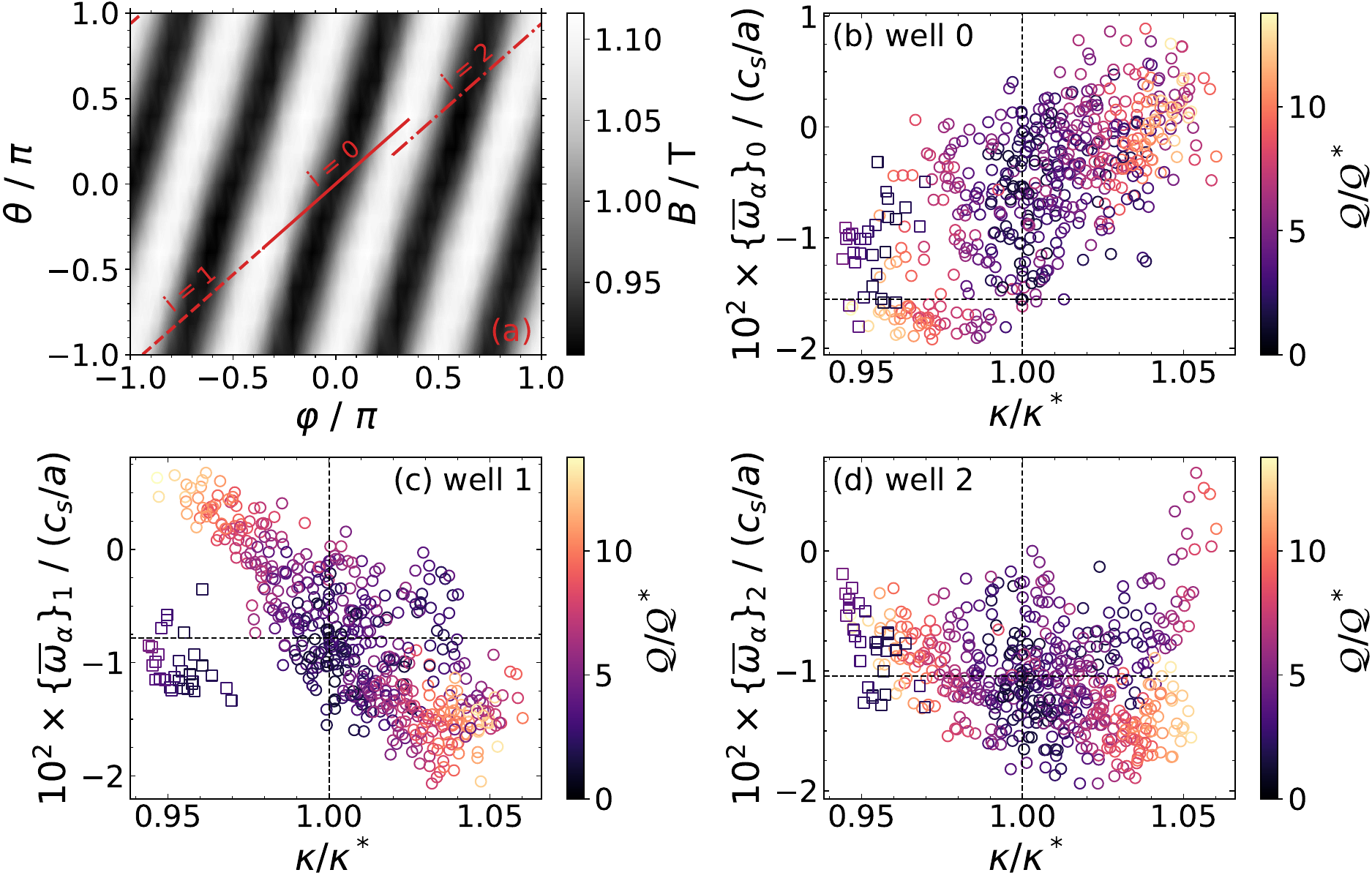}
		\caption{In panel (a) the magnetic field strength in the QHS configuration is plotted as a function of straight poloidal and toroidal field-line coordinates, and the characteristic $\alpha = 0$ field line is shown in red. The field line is partitioned into three separate trapping wells indexed by $i$. In panels (b)--(d) the velocity-space-averaged bounce-averaged precessional drift frequency is shown for trapping wells 0, 1, and 2 as a function of elongation, with the symmetry-breaking ratio shown in color. The black cross-hairs denote the QHS configuration. This shows how the particle drifts are affected by elongation across these three trapping wells. \label{fig:part_res}}
	\end{figure}

	Motivated by these observations, the impact of elongation on the velocity-space-averaged precessional drifts will be investigated in individual trapping wells instead of along the entire magnetic field line. This is done in Fig.~\ref{fig:part_res}, where three individual trapping wells are identified for each flux tube, with panel (a) showing a characteristic partitioning of these wells in the QHS configuration, labeling trapping wells as $i = 0$, $1$, or $2$. These three wells are selected because they span one complete poloidal rotation. In panels (b), (c), and (d), the velocity-space-averaged precessional drift for the electrons populating each trapping well is shown as a function of elongation, with the symmetry-breaking ratios for each configuration shown in color. Recall that $\{\overline{\omega}_{\alpha}\} < 0$ is destabilizing for TEMs while $\{\overline{\omega}_{\alpha}\} > 0$ is stabilizing.
	From Fig.~\ref{fig:part_res}(b), (c), and (d), the first thing to be observed is that the dependence of $\{\overline{\omega}_{\alpha}\}$ on elongation is qualitatively different in each trapping well, an observation that is consistent with the lack of any trend in growth rates with respect to $\{\overline{\omega}_{\alpha}\}$ when the latter is calculated along the entire flux-tube. The starkest comparison is between $\{\overline{\omega}_{\alpha}\}_0$ and $\{\overline{\omega}_{\alpha}\}_1$, shown in panels (b) and (c), respectively, with the subscript denoting the $i^{\mathrm{th}}$ trapping well. It can be observed that these wells exhibit opposite trends, where $\{\overline{\omega}_{\alpha}\}_0$ and $\{\overline{\omega}_{\alpha}\}_1$ are shifted towards stabilizing and destabilizing values, respectively, as elongation increases. However, if one considers the symmetry-breaking ratio $\mathcal{Q}/\mathcal{Q}^*$, then $\{\overline{\omega}_{\alpha}\}_0$ can be seen to shift toward stabilizing values at high elongation while preserving good quasi-helical symmetry in some configurations. Alternatively, the most stabilizing values of $\{\overline{\omega}_{\alpha}\}_1$ only appear in low-elongation configurations with severely broken quasi-helical symmetry. Similarly, in well 2, where the dependence on elongation is more complicated, the fact remains that more configurations with good quasi-helical symmetry, and $\{\overline{\omega}_{\alpha}\}_2$ shifted towards zero, are found at high elongation. Therefore, in HSX, if one is restricted to considering only configurations with relatively good quasi-helical symmetry, then increasing elongation will reduce the destabilizing drifts in some, but not all, magnetic trapping wells, thereby reducing linear growth rates without eliminating the TEMs. Similarly, the increase in the destabilizing drifts in well 1 with increasing elongation may contribute to the increase in growth rates at high elongation observed in Fig.~\ref{fig:gene_vs_kappa}, though this effect is difficult to distinguish from the effect of breaking the quasi-helical symmetry, which occurs simultaneously. Therefore, it remains unclear whether growth rates would continue to decrease with increasing elongation if the quasi-helical symmetry could be preserved to arbitrarily high elongation.
	
\begin{figure}
		\centering
		\includegraphics[width=.75\textwidth, keepaspectratio]{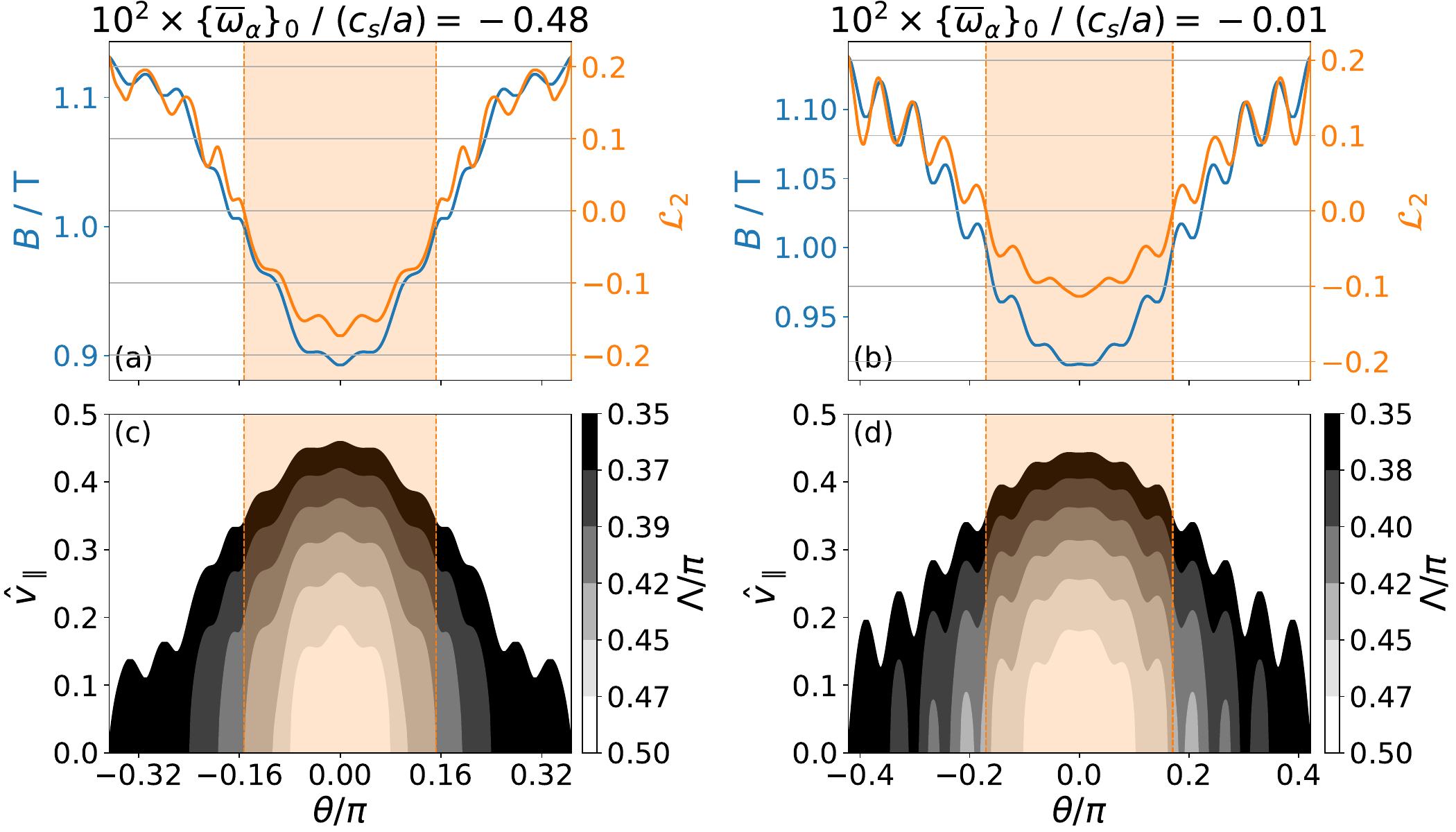}
		\caption{The magnetic trapping well crossing the outboard mid-plane is shown for two different configurations. In (a) and (b) the magnetic field strength and curvature drive are shown in blue and orange, respectively, each plotted as a function poloidal angle. In (c) and (d), the normalized-parallel-velocity contours are shown for the corresponding trapping wells above. The parallel velocity color bands indicate the pitching angle of the trapped electrons. In both top/bottom sets the region of destabilizing curvature is highlighted in orange. The trapping well on the right has higher elongation, demonstrating the flattening of the well that leads to the increase in $\{\overline{\omega}_{\alpha}\}$, as shown atop panels (a) and (b). \label{fig:trapping_wells}}
	\end{figure}

	To investigate the drift reversal in well 0 with increasing elongation, Fig.~\ref{fig:trapping_wells} shows the magnetic-field strength, field-line curvature $\mathcal{L}_2$, and normalized-parallel-particle velocity $\hat{v}_{\parallel} = v_{\parallel}/v$ as a function of poloidal angle in a set of characteristic trapping wells for two different configurations. In panels (a) and (b) the magnetic field strength is shown in blue and the curvature in orange, the latter of which determines the precessional drifts as described in Eq.~(\ref{eq:alpha_drift_gist_final}). Then, in panels (c) and (d), normalized-parallel-velocity contours are shown, with their dependence on pitch angle $\Lambda$ indicated by the various gray-scale bands. In both the top and bottom panels, the poloidal domain of the destabilizing curvature is highlighted in orange. Note that the configuration shown in panels (a) and (c) has $\kappa/\kappa^* = 0.989$ and that shown in (b) and (d) has $\kappa/\kappa^* = 1.023$.
	
	From Fig.~\ref{fig:trapping_wells} it can be observed that the trapping well in (b) and (d) is more extended in the poloidal domain. This can be attributed to an increase in the configuration's rotational transform, which results from an increase in its helical elongation (see Fig.~9 in Ref.~\cite{Gerard_2023}). One significant effect of this extension can be observed in panels (c) and (d), where the maximal pitch angle for which a particle's bounce point ($v_{\parallel} = 0$) occurs in a region with stabilizing curvature ($\mathcal{L}_2>0$) is higher in the high elongation configuration. This means the high elongation configuration has fewer trapped particles that spend their entire trajectory in regions of destabilizing curvature.
	
	Another effect of elongation is that it excites higher-order Boozer harmonics. This is observed in panels (a) and (b) of Fig.~\ref{fig:trapping_wells}, where the more elongated configuration has a larger magnetic-field ripple. Importantly, the symmetry-breaking ratio in these two configurations is nearly identical, with $\mathcal{Q}/\mathcal{Q}^* = 1.772$ and $1.784$ for the left and right configurations, respectively. This is possible because a significant fraction of the harmonics excited in the more elongated configuration are attributed to higher-order symmetry modes such as $(n,\,m) = $ $(8,\,2)$ and $(12,\,3)$. Analogous to a square well, the inclusion of these higher-order harmonics results in a flattening of the magnetic field strength towards the bottom of the trapping well. This flattening then allows for higher parallel velocities through the region in which $\mathcal{L}_2$ is negative, meaning the destabilizing curvature's contribution to the bounce-averaged precessional drift is minimized.
	
\begin{figure}
		\centering
		\includegraphics[width=.48\textwidth, keepaspectratio]{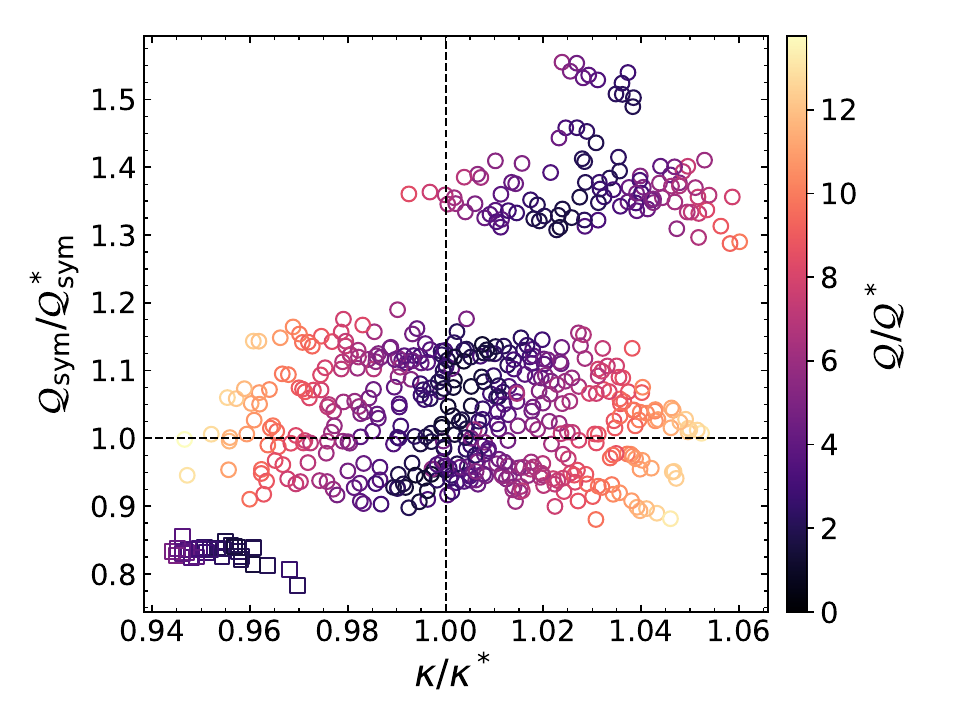}
		\caption{The higher-order symmetry metric is shown as a function of elongation, with the symmetry-breaking ratio shown in color. This shows that increasing elongation while preserving quasi-helical symmetry leads to an excitation of higher-order symmetry harmonics in the magnetic-field-strength spectrum. The black cross-hairs indicate the QHS configuration and the horizontal gaps in the data are explained in the text. \label{fig:high_sym}}
	\end{figure}

	To demonstrate this effect of elongation throughout the down-sampled database, a new metric intended to capture the presence of high-order symmetry modes is introduced. Similar to Eq.~(\ref{eq:symmetry}), which quantifies the symmetry breaking in the magnetic-field-strength spectrum, this new quantity is defined as
	\begin{equation}
		\mathcal{Q}_{\mathrm{sym}} = \frac{1}{B_{0,0}}\left( \sum_{j=2}^{J} B_{4j,j}^2 \right)^{1/2}, \label{eq:high_symmetry}
	\end{equation}
	where $J$ is the number of symmetry-preserving modes. To exclude the dominant symmetry mode $(n,m) = (4,1)$, the summation begins with $j=2$. For the QHS configuration $\mathcal{Q}_{\mathrm{sym}}^* = 3.923\times 10^{-3}$. Then, Fig.~\ref{fig:high_sym} shows $\mathcal{Q}_{\mathrm{sym}}/\mathcal{Q}_{\mathrm{sym}}^*$ as a function of elongation, with the symmetry-breaking ratio shown in color. The horizontal gaps observed in the data near $0.85 \lesssim \mathcal{Q}_{\mathrm{sym}}/\mathcal{Q}_{\mathrm{sym}}^* \lesssim 0.91$ and $1.15 \lesssim \mathcal{Q}_{\mathrm{sym}}/\mathcal{Q}_{\mathrm{sym}}^* \lesssim 1.35$ are due to the exclusion of configurations exhibiting strong magnetic resonances due to low-order rationals in their rotational transform. More interesting, however, is that if one considers configurations with good quasi-helical symmetry, indicated by low $\mathcal{Q}/\mathcal{Q}^*$, then increasing elongation results in an increase in the amplitudes of the higher-order symmetry modes, with amplitudes increasing by 50\% in the most elongated quasi-helically symmetric configurations.
	
	These shaping effects are subtle, but their influence on the particle drifts is profound, as exemplified by the near-reversal in $\{\overline{\omega}_{\alpha}\}$ shown at the top of Fig.~\ref{fig:trapping_wells}(a) and (b) and more broadly in Fig.~\ref{fig:part_res}(b). This then supports the conjecture made in Sec.~\ref{sec:selection}, where it was stated that TEM stability optimization could target the higher-order modes in the magnetic spectrum while preserving the overall quasi-helical symmetry. This is an important observation that can be leveraged in future optimization.
	
	\subsection{Impact of quasi-helical symmetry on TEM resonance}
	
\begin{figure}
		\centering
		\includegraphics[width=.48\textwidth, keepaspectratio]{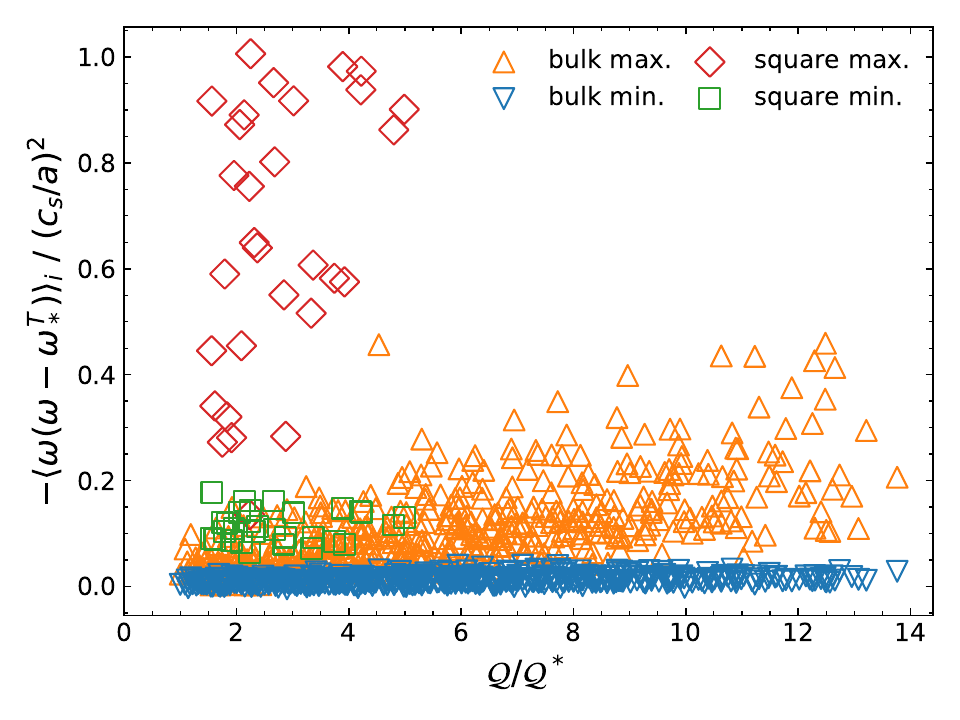}
		\caption{The maximum and minimum values of the velocity-space-averaged resonance operator, determined by calculating the resonance within all partitioned trapping wells along each flux tube, are shown as a function of the symmetry-breaking ratio. This shows that the resonance within the minimally resonant well does not change as the quasi-helical symmetry is broken, while the resonance of the maximally resonant well is increased. Therefore, a broken-symmetry configuration has an increased likelihood of producing a highly resonant trapping well, leading to greater instability over the entire flux tube. \label{fig:res_qhs}}
	\end{figure}

	The question, however, remains, why does such optimization require that quasi-helical symmetry be preserved? To address this, the partitioning of magnetic trapping wells demonstrated in Fig.~\ref{fig:part_res}(a) is applied to the entire flux-tube domain. For each configuration, $\langle\omega(\omega-\omega_*^T)\rangle_i$ is calculated in each partitioned trapping well, where $i$ indexes each well, and all radial wavenumbers are included. Then, the maximum and minimum of $-\langle\omega(\omega-\omega_*^T)\rangle_i$ in each flux-tube is shown in Fig.~\ref{fig:res_qhs} as a function of the symmetry-breaking ratio. The negative of the resonance operator is considered because maximal values of $-\langle\omega(\omega-\omega_*^T)\rangle_i$ indicate a maximally resonant and destabilizing trapping well. Note that the outlier configurations in Fig.~\ref{fig:resonance}, with $M_{\phi\times n} < -0.86$, are excluded because of their anomalously large $-\langle\omega(\omega-\omega_*^T)\rangle$. Moreover, due to the particularly large growth rates and resonance operator of the large-negative-squareness configurations, Fig.~\ref{fig:res_qhs} distinguishes these highly-shaped configurations from the bulk configurations, with the former shown as green squares and red diamonds for the minimum and maximum values, respectively, and the latter as upside-down blue triangles and right-side-up orange triangles for the minimum and maximum values, respectively.
	
	From Fig.~\ref{fig:res_qhs}, it can be observed that the minimally resonant well shows no dependence on the quasi-helical symmetry in either the bulk or square configurations, while the resonance in the maximally resonant well increases as the symmetry is broken. A likely explanation for this increase comes from the realization that breaking quasi-helical symmetry results in non-symmetric trapping wells that are confined to a field line within the domain of a symmetric trapping well. These non-symmetric trapping wells can then trap particles in regions with exclusively destabilizing curvature, leading to large destabilizing precessional drifts. Therefore, the inclusion of non-symmetric trapping wells can significantly increase the TEM resonance over a particular symmetry well.
	
\begin{figure}
		\centering
		\includegraphics[width=.48\textwidth, keepaspectratio]{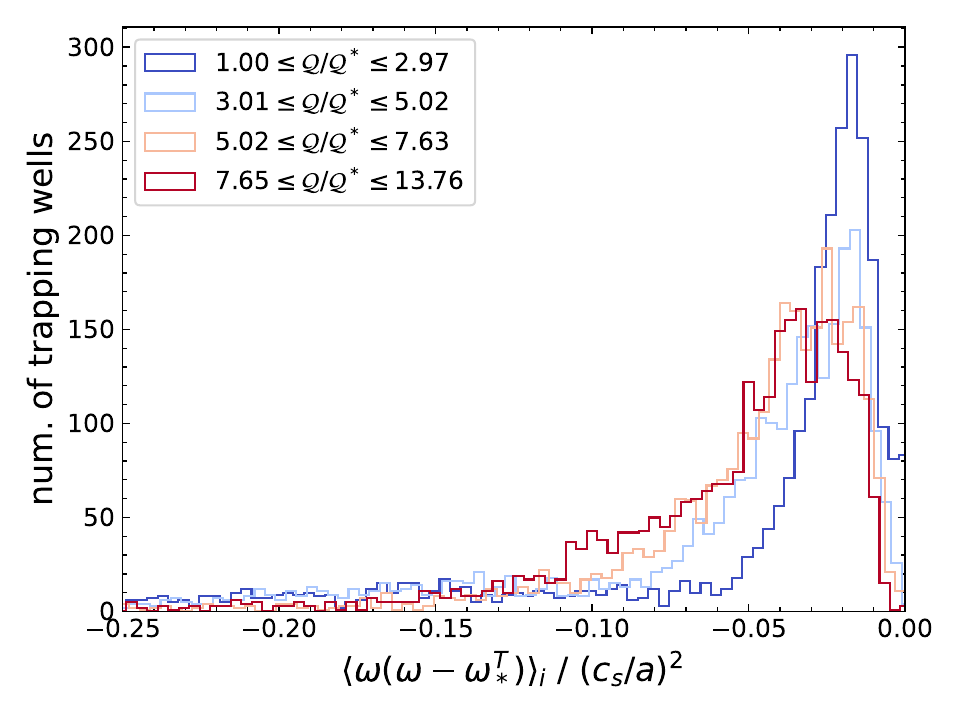}
		\caption{Four histograms are shown for the TEM resonance operator across all individual magnetic trapping wells. The data compiled in each histogram stems from configurations that span four different ranges of quasi-helical-symmetry ratios. This shows that breaking the quasi-helical symmetry broadens the distribution of trapping-well resonances to include a greater number of highly destabilizing trapping wells. \label{fig:histogram}}
	\end{figure}

	In addition to increasing the resonance in the maximally resonant trapping well, breaking quasi-helical symmetry also increases the prevalence of comparably resonant wells. This is demonstrated in Fig.~\ref{fig:histogram}, where four histograms of $\langle\omega(\omega-\omega_*^T)\rangle_i$ are shown, with each histogram comprised of data from configurations within four separate ranges of symmetry-breaking ratios. Note that these ranges in $\mathcal{Q}$ are not equal because the partitioning was done so that each set is comprised of trapping-well data from a nearly equal number of configurations. From the figure, it can be observed that the most quasi-helically symmetric configurations exhibit a sharp peak in their distribution of trapping-well resonances, with the majority of trapping-well resonances shifted toward zero. Alternatively, as more broken-symmetry configurations are considered, the distribution of trapping-well resonances broadens to include more destabilizing trapping wells.
	
	It is therefore concluded that a quasi-helical symmetry configuration will be less unstable relative to a broken-symmetry configuration because of the minimal variation in the underlying resonance across the individual trapping wells. Alternatively, in a broken-symmetry configuration, the inclusion of non-symmetric trapping wells will increase the likelihood that more strongly resonant wells will be accessible, leading to greater instability. This is in agreement with the observations made previously based on $E_\mathrm{A}$, namely that a reduction in $E_\mathrm{A}$ is observed with improved quasi-helical symmetry. Since the observed dependence of $E_\mathrm{A}$ on quasi-helical symmetry is best explained by the dependence of $E_\mathrm{A}$ on the bounce-averaged radial particle drifts $\omega_{\psi}$, these results provide some evidence for the more general claim that quasi-omnigeneity might limit TEM growth rates by reducing available energy in the system.
	
	\section{Conclusions \label{sec:conclusions}}
	
	Using an HSX coil-current database \cite{Gerard_2023}, a set of 563 HSX equilibria has been compared against the QHS configuration to identify trends between macroscopic shaping parameters and TEM stability. It has been found that the lowest TEM growth rates occur in helically elongated geometries that preserve the quasi-helical symmetry of the QHS configuration. Moreover, the reduction in growth rates with improved quasi-helical symmetry is shown to be consistent with predictions from available-energy calculations. However, both $E_\mathrm{A}$ and quasi-helical symmetry fail to predict the reduction with respect to elongation.
	
	Alternatively, the velocity-space-averaged resonance operator has been shown to accurately capture the reduction in TEM growth rates. However, this resonance calculation requires knowledge of the drift-wave frequency and eigenmode structure, precluding this quantity as an easy-to-evaluate TEM metric. The quantity does demonstrate utility in revealing the physical mechanisms that contribute to the reduction in growth rates. For example, it has been found that the reduction in growth rates with increasing elongation occurs by reducing the TEM resonance without supporting a maximum-$\mathcal{J}$ flux surface. Moreover, breaking the quasi-helical symmetry is shown to broaden the distribution of trapping-well resonances to include a greater number of highly destabilizing trapping wells.
	
	Regarding general trends towards optimization, it has been found that the elongated configurations of HSX produce lower TEM growth rates because of the extension of the trapping wells in the poloidal domain as well as the inclusion of high-order symmetry harmonics in their Boozer spectrum. The latter suggests that if a configuration has been optimized for quasi-helical symmetry, then further optimization to increase the amplitude of high-order symmetry harmonics may be beneficial for reducing TEM growth rates. However, it is not clear, at present, how such an optimization would be performed since the sign of a particular symmetry mode will determine whether that mode steepens or flattens a magnetic well, and Eq.~(\ref{eq:high_symmetry}) does not preserve such information. Therefore, it may be easier to target the excitation of these modes by simply targeting increased plasma elongation, with a strong penalty for configurations that break the quasi-helical symmetry. Notably, consistency between the quasi-helical symmetry metric and $E_\mathrm{A}$, as seen when comparing Figs.~\ref{fig:gene_vs_ae} and \ref{fig:gene_vs_qhs}, suggests that a similar approach could be to target both increased elongation and reduced $E_\mathrm{A}$ simultaneously.
	
	The next step is to perform nonlinear simulations to verify key findings, which will be reported in a future publication. Regarding the experimental validation of these observations, plans are underway to measure density and temperature fluctuations and to calculate heat fluxes from power balance analysis in HSX geometries identified from the coil-current database.
	
	\begin{acknowledgments}
		The authors thank H.H.~Oliveira~Miller and S.~Stewart for helpful discussions. Support was received through U.S. DOE Grant No.~DE-SC0020990. Computing time was provided through the National Energy Research Scientific Computing Center, a DOE Office of Science User Facility, Allocation No.~m3586, as well as the University of Wisconsin-Madison's Center for High Throughput Computing.
	\end{acknowledgments}
	
	\appendix
	
	\section{Bounce-Averaged Drifts from GIST \label{appdx:drifts}}
	
	To calculate the bounce-averaged-drift frequencies for species $s$ first consider that
	\begin{align}
		\overline{\omega}_{\alpha} &= k_{\alpha} \overline{ \mathbf{v}_\mathrm{ds} \cdot \nabla\alpha }    \label{eq:alpha_drift} \\
		\overline{\omega}_{\psi} &= k_{\psi} \overline{\mathbf{v}_\mathrm{ds} \cdot \nabla\psi}, \label{eq:psi_drift}
	\end{align}   	 
	with the over-line defining a bounce average
	\begin{equation}
		\overline{h} = \frac{\int d\ell h/v_{\parallel}}{\int d\ell/v_{\parallel}}. \label{eq:bounce-average}
	\end{equation}
	In this paper, all such bounce averages are performed using the general trapezoidal rule described in Sec.~III.C of Ref.~\cite{Mackenbach_2023_PoP}. The local magnetic drift is expressed as
	\begin{equation}
		\mathbf{v}_\mathrm{ds} = \frac{v_{\perp}^2}{2\Omega} \left( \frac{\mathbf{b} \times \nabla B}{B} \right) + \frac{v_{\parallel}^2}{\Omega} \left( \mathbf{b} \times \boldsymbol{\kappa} \right), \label{eq:drift_velocity}
	\end{equation}
	where $\Omega = ZeB/m$ is the cyclotron frequency.
	
	A field-aligned coordinate system is defined in Sec.~V of Ref.~\cite{xanthopoulos_geometry_2009} as
	\begin{align}
		u^1 &= \sqrt{s} \\
		u^2 &= \sqrt{s_0} \left[q(x^1)(\theta - \theta_k) - \varphi\right] \\
		u^3 &= \theta - \theta_k,
	\end{align}
	with $u^i = \{x,\, y,\, z\}$ for index $i=1$, $2$, and $3$, $\theta$ and $\varphi$ are the poloidal and toroidal coordinates in a straight field-line coordinate system, $\theta_k$ is the so-called ballooning angle \cite{beer_1995}, and the following normalizations are used
	\begin{align}
		\hat{B} &= \frac{B}{B_\mathrm{ref}}, &
		s &= \frac{\psi}{\psi_\mathrm{edge}}, &
		\hat{\nabla} &= a\nabla,
	\end{align}
	with $\psi_\mathrm{edge} = a^2B_\mathrm{ref}/2$. Then, defining
	\begin{align}
		\mathcal{L}_i &= \frac{1}{\hat{B}} \hat{\mathbf{b}}\times\hat{\nabla}\hat{B} \cdot \hat{\nabla}u^i \\
		\mathcal{K}_i &= \hat{\mathbf{b}} \times \hat{\boldsymbol{\kappa}} \cdot \hat{\nabla}u^i,
	\end{align}
	with $i=1$ or $2$, $\hat{\boldsymbol{\kappa}}=a\boldsymbol{\kappa}$, and $\hat{\mathbf{b}} = \hat{\mathbf{B}}/\hat{B}$, the differential curvature operators can be expressed
	\begin{equation}
		\left( \frac{\mathbf{b} \times \nabla B}{B} \right) \cdot \nabla = \frac{1}{a^2} \left( \mathcal{L}_1\partial_1 + \mathcal{L}_2\partial_2 \right), \label{eq:gradB_diff}
	\end{equation}
	and
	\begin{equation}
		\left( \mathbf{b} \times \boldsymbol{\kappa} \right) \cdot \nabla = \frac{1}{a^2} \left( \mathcal{K}_1 \partial_1 + \mathcal{K}_2\partial_2 \right), \label{eq:curve_diff}
	\end{equation}
	with $\partial_i = \partial/\partial u^i$. Therefore, one finds
	\begin{equation}
		\mathbf{v}_\mathrm{ds} \cdot \nabla = \frac{v_{\perp}^2}{2\Omega}\frac{1}{a^2} \left( \mathcal{L}_1\partial_1 + \mathcal{L}_2\partial_2 \right) + \frac{v_{\parallel}^2}{\Omega} \frac{1}{a^2} \left( \mathcal{K}_1\partial_1 + \mathcal{K}_2\partial_2 \right). \label{eq:direc_deriv}
	\end{equation}
	Rewriting this operator in terms of $\lambda = \mu B_{\mathrm{ref}}/E$, one obtains
	\begin{equation}
		\mathbf{v}_\mathrm{ds} \cdot \nabla = \frac{E}{2Ze\psi_{\mathrm{edge}}} \left[ \lambda \left( \mathcal{L}_1\partial_1 + \mathcal{L}_2\partial_2 \right) + 2\frac{1-\lambda\hat{B}}{\hat{B}} \left( \mathcal{K}_1\partial_1 + \mathcal{K}_2\partial_2 \right) \right]. \label{eq:direc_deriv_norm}
	\end{equation}
	Using $\psi = \psi_\mathrm{edge}x^2$ and $\alpha = y/\sqrt{s_0}$ (this is the definition of $\alpha$ as in Ref.~\cite{helander_theory_2014}, which differs by a factor of $q$ from that in Ref.~\cite{xanthopoulos_geometry_2009}), it is found that
	\begin{align}
		\mathbf{v}_\mathrm{ds} \cdot \nabla\alpha &= \frac{E}{2Ze\psi_\mathrm{edge}\sqrt{s_0}} \left( \lambda\mathcal{L}_2 + 2\frac{1-\lambda\hat{B}}{\hat{B}} \mathcal{K}_2 \right) \label{eq:drift_gist} \\
		\mathbf{v}_\mathrm{ds} \cdot \nabla\psi &= \frac{E\sqrt{s_0}}{Ze} \left( \lambda\mathcal{L}_1 + 2\frac{1 - \lambda\hat{B}}{\hat{B}} \mathcal{K}_1 \right). \label{eq:drift_psi}
	\end{align}
	
	Once again, from Ref.~\cite{xanthopoulos_geometry_2009} it is known that $\mathcal{L}_1 = \mathcal{K}_1$ and
	\begin{equation}
		\mathcal{K}_2 = \mathcal{L}_2 - \frac{\hat{p}^{\prime}(s_0)}{2\hat{B}}, \label{eq:curve_trans}
	\end{equation}
	with
	\begin{equation}
		\hat{p}^{\prime}(s_0) = -\frac{a^4\sqrt{s_0}}{\psi_\mathrm{edge}^2} \frac{dp}{ds}\bigg\rvert_{s_0}.
	\end{equation}
	Therefore, Eq.~(\ref{eq:drift_gist})--(\ref{eq:drift_psi}) can be expressed
	\begin{align}
		\mathbf{v}_\mathrm{ds} \cdot \nabla\alpha &= \frac{E}{2Ze \psi_\mathrm{edge} \sqrt{s_0}} \left[ \left( \lambda + 2 \frac{1-\lambda\hat{B}}{\hat{B}} \right) \mathcal{L}_2 - \hat{p}^{\prime} \left( \frac{1-\lambda\hat{B}}{\hat{B}^2} \right) \right]. \label{eq:drift_gist_alf} \\
		\mathbf{v}_\mathrm{ds} \cdot \nabla\psi &= \frac{E \sqrt{s_0}}{Ze} \left( \lambda + 2\frac{1 - \lambda\hat{B}}{\hat{B}} \right) \mathcal{L}_1. \label{eq:drift_gist_psi}
	\end{align}
	The normalized drift velocities in the binormal and radial directions are defined as
	\begin{align}
		\hat{v}_\mathrm{ds,y} &= \left( \lambda + 2 \frac{1-\lambda\hat{B}}{\hat{B}} \right) \mathcal{L}_2 - \hat{p}^{\prime} \left( \frac{1-\lambda\hat{B}}{\hat{B}^2} \right) \label{eq:alpha_drift_gist_final} \\
		\hat{v}_\mathrm{ds,x} &= \left( \lambda + 2\frac{1 - \lambda\hat{B}}{\hat{B}} \right) \mathcal{L}_1, \label{eq:psi_drift_gist_final}
	\end{align}
	respectively. Then, taking a bounce average of Eqs.~(\ref{eq:drift_gist_alf})--(\ref{eq:drift_gist_psi}) and scaling each by the corresponding wavenumber, normalized as shown in Eq.~(\ref{eq:gene_normalizations}), one finds
	\begin{align}
		\overline{\omega}_{\alpha} &= \left( \frac{\xi}{\rho_*} \right) \hat{\overline{\omega}}_\mathrm{y} \\
		\overline{\omega}_{\psi} &= q_0^2\left( \frac{\xi}{\rho_*} \right) \hat{\overline{\omega}}_\mathrm{x}
	\end{align}
	with $\hat{\overline{\omega}}_\mathrm{y} = \varepsilon\hat{k}_\mathrm{y}\hat{\overline{v}}_\mathrm{ds,y}$ and $\hat{\overline{\omega}}_\mathrm{x} = \varepsilon\hat{k}_\mathrm{x}\hat{\overline{v}}_\mathrm{ds,x}$.
	%
	
	The equilibria considered throughout this paper are all vacuum field configurations. As such, $\hat{p}^{\prime} = 0$ and $\mathcal{L}_2$ is identified, from Eq.~(\ref{eq:alpha_drift_gist_final}), as the curvature drive for the precessional drift. Note, $-\mathcal{L}_1$ and $\mathcal{L}_2$ are provided as a function of $z$ in the GIST output.
	
	\section{Cross-Phase Histograms \label{appdx:cross_phase}}
	
	To reduce the cross-phase histograms to a scalar, the cross-phase Gaussian weighting
	\begin{equation}
		w\left(\vartheta\right) = \exp{\left[ -N \left( 1 - \frac{2\vartheta}{\pi} \right)^2 \right]} \label{eq:guassian}
	\end{equation}
	is defined, where $\vartheta$ is the complex-phase angle and $N = 4\ln(10)$. This choice in $N$ is made because a destabilizing cross-phase for either trapped or passing particles requires that $\vartheta \in (0, \pi)$. When $\vartheta$ is outside of this range, $w$ is at least four orders of magnitude below its peak value.
	
	Labeling the cross-phase distribution as $D_{x}(\vartheta)$, where $x = t$ or $p$ for trapped and passing particles, respectively, the distribution average can be defined as
	\begin{equation}
		D^{\prime}_x = \frac{\int_{-\pi/2}^{3\pi/2} D_x(\vartheta) w(\vartheta) d\vartheta}{\int_{-\pi/2}^{3\pi/2} w(\vartheta) d\vartheta}. \label{eq:distribution}
	\end{equation}
	Note that the limits of integration are selected so that the integration domain has width $2\pi$ centered about $\pi/2$. Finally, these phase-space averaged distributions can be combined as the scalar
	\begin{equation}
		M_{\phi\times n} = \frac{D^{\prime}_t - D^{\prime}_p}{D^{\prime}_t + D^{\prime}_p}, \label{eq:mode_scalar}
	\end{equation}
	which has the desired effect of being $-1$ for a UI cross-phase signature, $+1$ for TEM, and $\approx 0$ for a mixed mode.
	
	\bibliographystyle{apsrev4-2}
	\bibliography{./paper_bib}
	
\end{document}